# A Conversation with Ingram Olkin

**Allan R. Sampson**

*Abstract.* Ingram Olkin was born on July 23, 1924 in Waterbury, Connecticut. His family moved to New York in 1934 and he graduated from DeWitt Clinton High School in 1941. He served three years in the Air Force during World War II and obtained a B.S. in mathematics at the City College of New York in 1947. After receiving an M.A. in mathematical statistics from Columbia in 1949, he completed his graduate studies in the Department of Statistics at the University of North Carolina in 1951. His dissertation was written under the direction of S. N. Roy and Harold Hotelling. He joined the Department of Mathematics at Michigan State University in 1951 as an Assistant Professor, subsequently being promoted to Professor. In 1960, he took a position as Chair of the Department of Statistics at the University of Minnesota. He moved to Stanford University in 1961 to take a joint position as Professor of Statistics and Professor of Education; he was also Chair of the Department of Statistics from 1973–1976. In 2007, Ingram became Professor Emeritus.

Ingram was Editor of the *Annals of Mathematical Statistics* (1971–1972) and served as the first editor of the *Annals of Statistics* from 1972–1974. He was a primary force in the founding of the *Journal of Educational Statistics*, for which he was also Associate Editor during 1977–1985. In 1984, he was President of the Institute of Mathematical Statistics. Among his many professional activities, he has served as Chair of the Committee of Presidents of Statistical Societies (COPSS), Chair of the Committee on Applied and Theoretical Statistics of the National Research Council, Chair of the Management Board of the American Education Research Association, and as Trustee for the National Institute of Statistical Sciences.

He has been honored by the American Statistical Association (ASA) with a Wilks Medal (1992) and a Founder's Award (1992). The American Psychological Association gave him a Lifetime Contribution Award (1997) and he was elected to the National Academy of Education in 2005. He received the COPSS Elizabeth L. Scott Award in 1998 and delivered the R. A. Fisher Lecture in 2000. In 2003, the City University of New York gave him a Townsend Harris Medal.

An author of 5 books, an editor of 10 books, and an author of more than 200 publications, Ingram has made major contributions to statistics and education. His research has focused on multivariate analysis, majorization and inequalities, distribution theory, and meta-analysis. A volume in celebration of Ingram's 65th birthday contains a brief biography and an interview [Gleser, Perlman, Press and Sampson (1989)]. Ingram was chosen in 1997 to participate in the American Statistical Association Distinguished Statistician Video Series and a videotaped conversation and a lecture (Olkin, 1997) are available from the ASA (1997, DS041, DS042).

*Allan R. Sampson is Professor, Department of Statistics, Cathedral of Learning 2701, University of Pittsburgh, Pittsburgh, Pennsylvania 15260, USA e-mail: asampson@stat.pitt.edu.*

This conversation took place on December 9, 2005 at the Renaissance Hotel in Washington, DC.

## INTRODUCTION AND EARLY YEARS

**Sampson:** Ingram, we've been trying to get together to do this conversation for quite a while. It's difficult to find any gap in your extremely hectic travel schedule. You're perhaps more involved in statistics now than you were when you began your career 55 years ago. And the question I want to start with is: what is it about statistics that's still so compelling to you?

**Olkin:** That's an interesting question. Statistics has a role in so many different applications, and what I always find exciting is the fact that you're confronted with a new discipline and a new set of people, and they bring different and interesting scientific questions in which the statistician can participate. Just to illustrate, I was most recently asked to discuss some aspects of what is called "the value of a statistical life" with the Environmental Protection Agency. This is a fascinating problem because it covers very large areas of the environment, and what's interesting is the people working on the projects are economists, not statisticians, so it brings me into contact with a whole new area.

**Sampson:** You've been involved with these various forms of applied problems for a long time. But what keeps you still so fired-up?

**Olkin:** This question about the "fire" is one I have trouble answering.

**Sampson:** It may precede your beginnings in statistics—perhaps something in your upbringing.

**Olkin:** I suspect that's true. My mother had fire until she was 98.

**Sampson:** Amen!

**Olkin:** The "fire" was mostly addressed to me.

And I think there may be some genetics because my daughters have a certain amount of that transmitted.

**Sampson:** When you were a child, were you as intense and as passionate in whatever you were doing then as you are now?

**Olkin:** Let me put it this way. I was born in Waterbury, Connecticut and I lived there until age 10. I went to a public school that had very little heat, and when you misbehaved they sent you to sit in the cloakroom, which was cold. What I can tell you is my mother was a constant visitor to the school because I was in the cloakroom an inordinate amount of time. So, I think I was a very active kid at that time.

**Sampson:** It sounds more than active, but perhaps rebellious.

**Olkin:** I don't know. But I do know, and I can't tell you what grade, the teacher taped my mouth.

**Sampson:** Some of your colleagues now, I suspect, wish they knew that teacher's secret!

**[Laughter]**

**Olkin:** Now, of course, this was abuse under current definitions. But, it didn't bother me at the time. I think I've always been active and interested in different things. I think that certainly is a characteristic.

**Sampson:** Did you have a lot of interests as a child?

**Olkin:** Yes I did. And I think I was sort of good in a lot of different areas and so I was interested in sports, I collected stamps, I went to theaters, music, and so on. I really availed myself.

**Sampson:** Some of that sounds like it occurred when you were older.

**Olkin:** Theater and music occurred during my early teens after I moved to New York. There were not that many outside interests available in Waterbury, Connecticut. The only place that was available was a library, and my mother took me to the library on many, many occasions. It was a ritual. That's an interesting point because my mother was an immigrant who really didn't have much education.

**Sampson:** Where was your mother from?

**Olkin:** My mother came from Warsaw, my father came from Vilnius, but he was in Warsaw at the time, and so they both came over. But my mother knew that books were good. She dragged me to the library, but she didn't have to drag very much. It was just nice to go there every Saturday morning. By the time I got to New York, I was independent—even though I was only 10 years old. The thing about New York is that children were totally independent.

**Sampson:** You lived in the Bronx?

**Olkin:** I lived in the Bronx on Arthur Avenue and 179th Street.

**Sampson:** You finished elementary school in New York City?

**Olkin:** I finished elementary school in New York City and then went to middle school followed by Dewitt Clinton High School. That was a very good high school at the time. Dewitt Clinton had an annex and it was that annex which became the Bronx High School of Science. They had superior teachers.

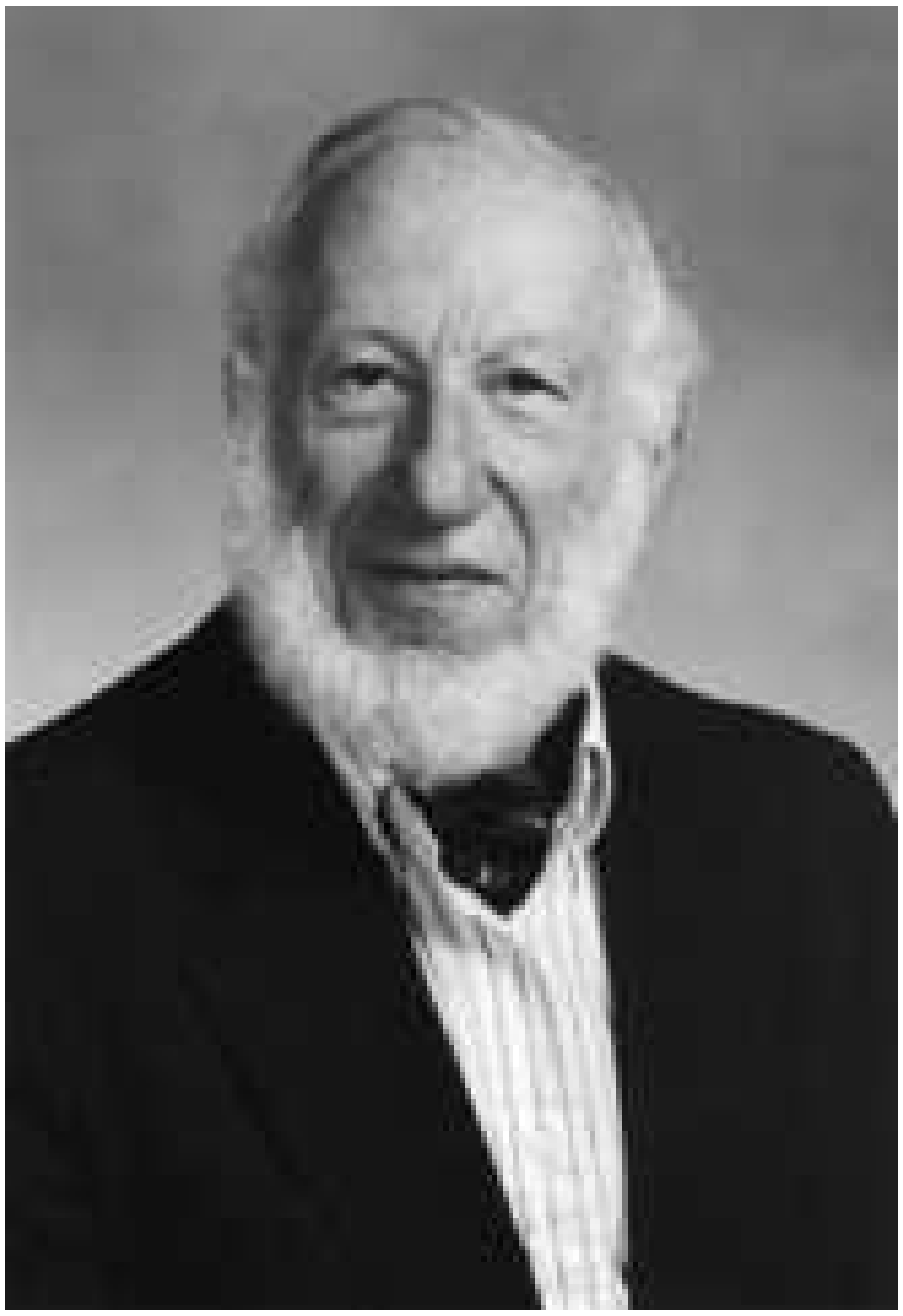

FIG. 1. *Ingram Olkin, 2001.*

The math teacher actually published in *The American Mathematical Monthly*. There was a math club in high school, and there was music. And at that time I played trombone, and I was in the orchestra.

**Sampson:** You were an only child?

**Olkin:** I was the only child. My mother actually lost several children, some in miscarriage, and one child at a year old. I was really somewhat of a last chance because they were getting older by the time I was born. My mother was 36. At that time, giving birth at that age was probably already precarious,

**Sampson:** Your dad was in the jewelry business in Waterbury?

**Olkin:** Yes. In those years, it was very customary for immigrants to have a relative or friend in the United States who would sort of set the stage for them. My father had a very close friend who had a jewelry business in Waterbury, and my father was a jeweler in Europe. So he had a job ready-made in Waterbury and that's how we got there. I was born in 1924 and this is pre-depression. When I was a child, my father actually was out of work for quite a bit of time though my mother didn't know it. He never told her, and somehow we managed. But it was clear that Waterbury was somewhat of a dead-end in the jewelry business, which is one of the first to go during any depression in any case. But there was a second factor of why we moved to New York. By this time, I was getting close to 10 years old, and the question was what would the future bring in terms of college? Connecticut was not known at that time—we're talking about the 30s—for lots of good state universities. And, in fact, I had a cousin who was teaching at City College of New York, and the general view was that City College was the place where I could get an education. We moved to New York in 1934, and it was clear that I would go to City College because that was the only place open in the immediate post-depression era.

## NEW YORK

**Sampson:** Did your father eventually find jewelry work in New York?

**Olkin:** Yes. In the jewelry trade, they'll have one big store with little stalls. And he had a stall on Canal Street which had a jewelry district and one on 47th Street. Throughout his life he continued to work in the jewelry trade.

**Sampson:** From the way you describe it, it sounds like your mother was more influential in your upbringing than your father. Is that a fair statement?

**Olkin:** There's no question about that. My mother was a very strong woman. She had firm ideas, and she was also a very active type. My father was a very gentle kind person, whom everyone in the family and the extended family thought was great. Nobody ever had an unkind word about my father. My mother would generate different reactions from different people, but she was a very positive force.

**Sampson:** Did she have any influence in your going into the mathematical sciences?

**Olkin:** No, I have to say in thinking about my upbringing and the upbringing of our children, there's a distinct difference. My mother did not know what I was doing in college or in high school. But she believed that whatever I did was great, so I had complete full support for whatever I wanted to do. She had faith that no matter what I would do, it would be fine. And now I think of our own actions with our children. My daughter would come home and say, "Ok, I'm taking math." I would say, "Well, what math?" She would say, "Algebra." I'd say, "What are you studying in algebra?" and so on. There's something very positive about unequivocal support that really is very healthy.

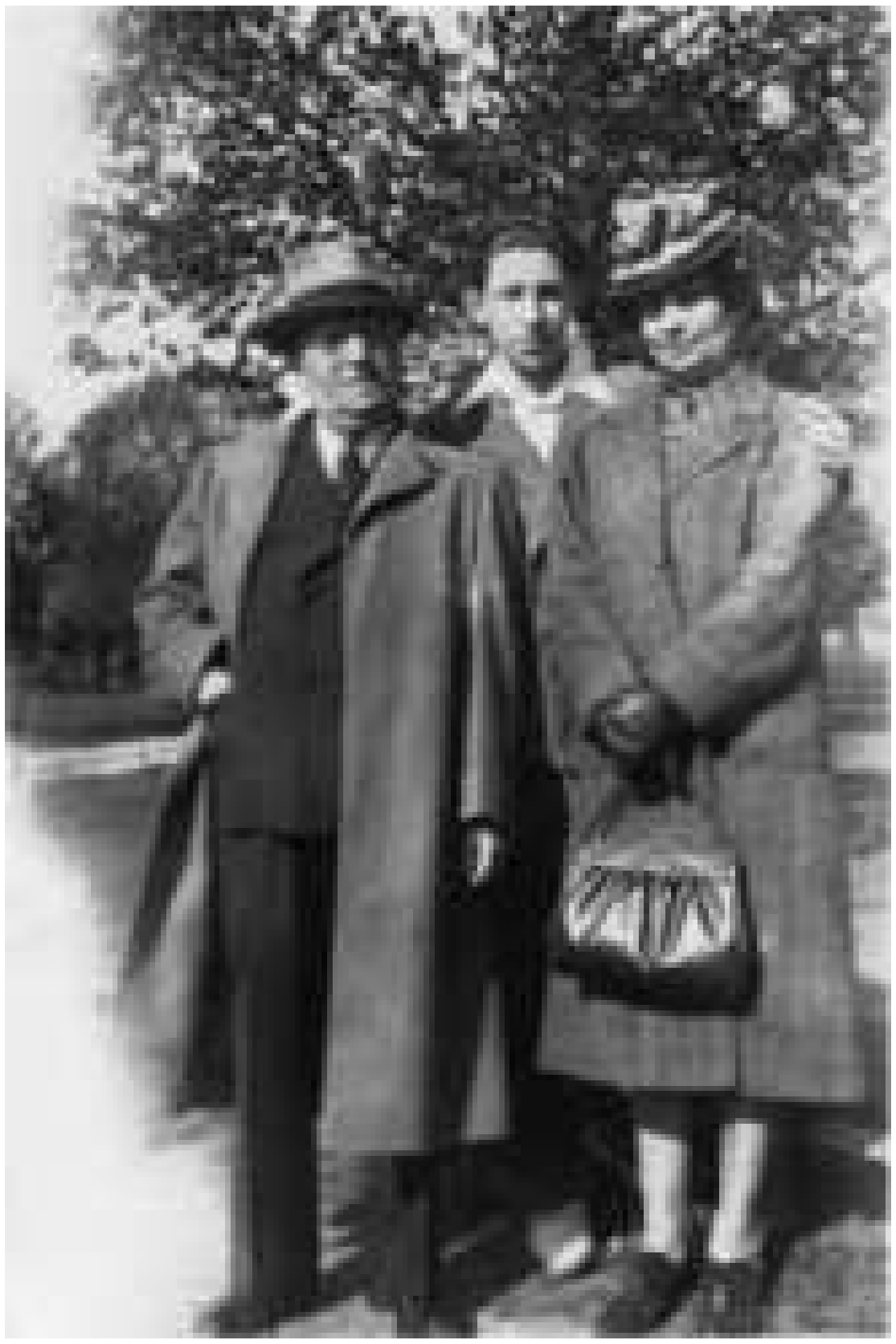

Fig. 2. *Ingram Olkin and his parents, Julius and Karola Olkin, 1940s.*

Early on in every Jewish family, there was always the question of whether you're going to be a doctor. And it became very clear at an early age that I did not like anything connected with blood. Thus I never had to go through the period of "my son, the doctor." My mother always said, "Well, he doesn't like medicine" and that left me out of that.

**Sampson:** As an aside, I am interested in your family's involvement with Judaism.

**Olkin:** My parents were what you would call mild orthodox. Our house was kosher, but my father, when times were bad, did go to work on Saturday. We always observed all the holidays, and he did go to the synagogue on Saturdays a lot of the time. And I would accompany him. I was certainly Bar Mitzvahed, and married in the synagogue, and our house was what you might call medium orthodox.

**Sampson:** Let me come back to your high school again. I know you once told me that when they asked you in your high school year book what you intended to be, you said, "a statistician."

**Olkin:** That is correct. It's clear now, to both of us, that I didn't know anything of what being a statistician meant.

**Sampson:** Are you any wiser 65 years later?!

**[Laughter]**

**Olkin:** That's a good question! I have a feeling that the math club had lots of different problems, and I suspect that there were some problems in statistics that we went through that must have captured my imagination. But I cannot now reproduce how I ever chose the term statistician. I may be the only one in the profession who in high school said that they wanted to be a statistician, and it came true. Now, it's interesting to go back to the high school yearbook, which I did. Several people were well known. For example, James Baldwin, the author, was at Dewitt Clinton High School. I checked to see what he said he wanted to be. He said he wanted to be a writer. There were several others whose stated goals in high school came true in later life. Somehow, there may have been some germs of what people thought they might be interested in, and I suspect that statistician just meant some part of mathematics without my really knowing what it was.

**Sampson:** You started college at CCNY and you had to take a break for the war, before returning to CCNY?

**Olkin:** I graduated Dewitt Clinton High School in 1941. I started in CCNY and then in late 1942, the government—and I don't know how this came to me—said they were interested in students in mathematics, physics and possibly engineering, going into radar, meteorology, and languages. There was something called the ASTP, the Armed Services Training Program. Even though math majors were deferred, I thought that eventually I would be drafted. Everybody was drafted at the time, and so I enlisted in the Meteorology Program. I was inducted in February of 1943. I was away from 1943 to 1946, and then I was discharged. My discharge was in California, but I returned to New York and finished City College in 1947.

After that I went to Columbia for a master's degree which was a one-year program. I started in the summer of 1947, and then finished the following year. The summer of 1947 was one of the first post World War II classes and it was an exciting time at Columbia. Many of the students who were there became close life-long friends including Bob Bechhofer, Milton Sobel and Rosedith Sitgreaves. Also that was a time when Hotelling had moved from Columbia to Chapel Hill (in 1946). While at Columbia, I applied to Chapel Hill for the Ph.D. program and for a fellowship, and we then moved to Chapel Hill in 1948.

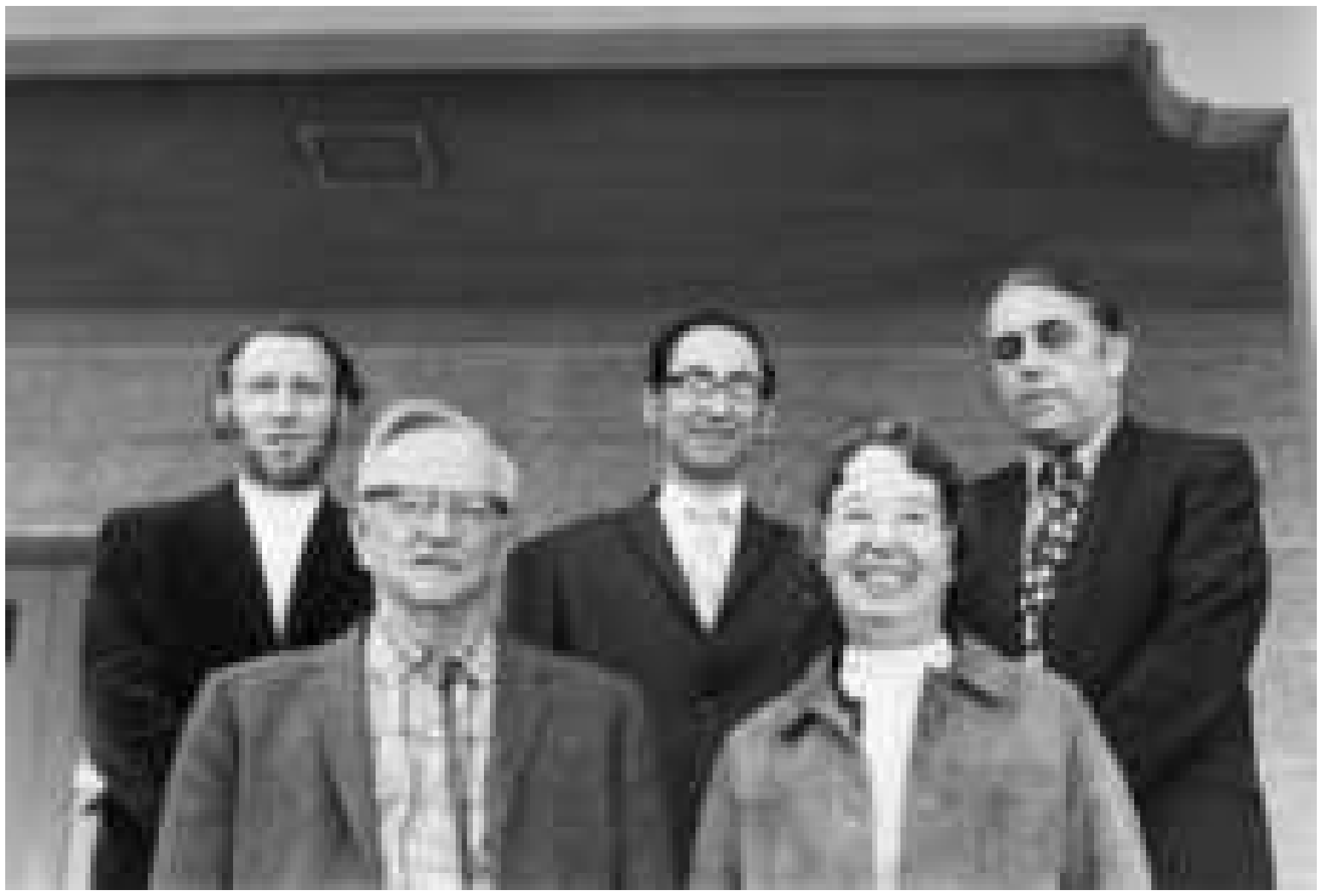

Fig. 3. *Mr. and Mrs. Selby Robinson (bottom) with his former City College of New York students: Ingram Olkin, Herman Chernoff, Herb Solomon (top), Stanford, 1971.*

## GRADUATE SCHOOL AT U. NORTH CAROLINA

**Sampson:** Tell me a bit about your experiences at Chapel Hill.

**Olkin:** I think one has to think somewhat in terms of the history of statistics. Nineteen forty-six was the time when statistics was really beginning in several places. Columbia had a department in 1946, with Abraham Wald, Jack Wolfowitz, Ted Anderson and Howard Levene, and I think Henry Scheffé was there at the time. Princeton did not have a Department of Statistics, but it had Sam Wilks and John Tukey. Berkeley, of course, had Jerzy Neyman and Erich Lehmann. They did not have a department in 1946—that came later. But they had a statistical laboratory and many eminent faculty. Chicago was another place; Iowa State; North Carolina State started; but Chapel Hill was a galaxy of faculty. Harold Hotelling was the leader. Herb Robbins was there, Wassily Hoeffding was there, as were R. C. Bose, S. N. Roy and P. L. Hsu, although Hsu was on leave and it was not clear whether he was coming back. William Cochran was at Raleigh, but came to Chapel Hill on many occasions. Bill Madow was there.

**Sampson:** Gertrude Cox?

**Olkin:** Gertrude Cox was the head of the entire Institute, and she was housed more at Raleigh. In any case, Chapel Hill was a phenomenal place. The students were great. At that time, India had supported a lot of students to study in the United States. This was really the "heyday" of Indian statistics in the United States. Raj Bahadur was there as a student. S. S. Shrikhande, Gopinath Kallianpur and D. N. Nanda were students, and later on, K. C. S. Pillai, Shanti Gupta, Ram Gnandesikan and Govind Mudholkar. I can go on, but I don't remember them all.

**Sampson:** Sudhish Ghryre was someone that you worked with?

**Olkin:** He came a year later. But he was there. It was really a first rate group. And then there were American students: Ralph Bradley was a student, Meyer Dwass was there, Joan Rosenblatt, Morris Skibinsky, Sutton Monro, and many others. There was a lot of camaraderie among the students, and there were very few barriers between faculty and students. Part of it was that the faculty was young.

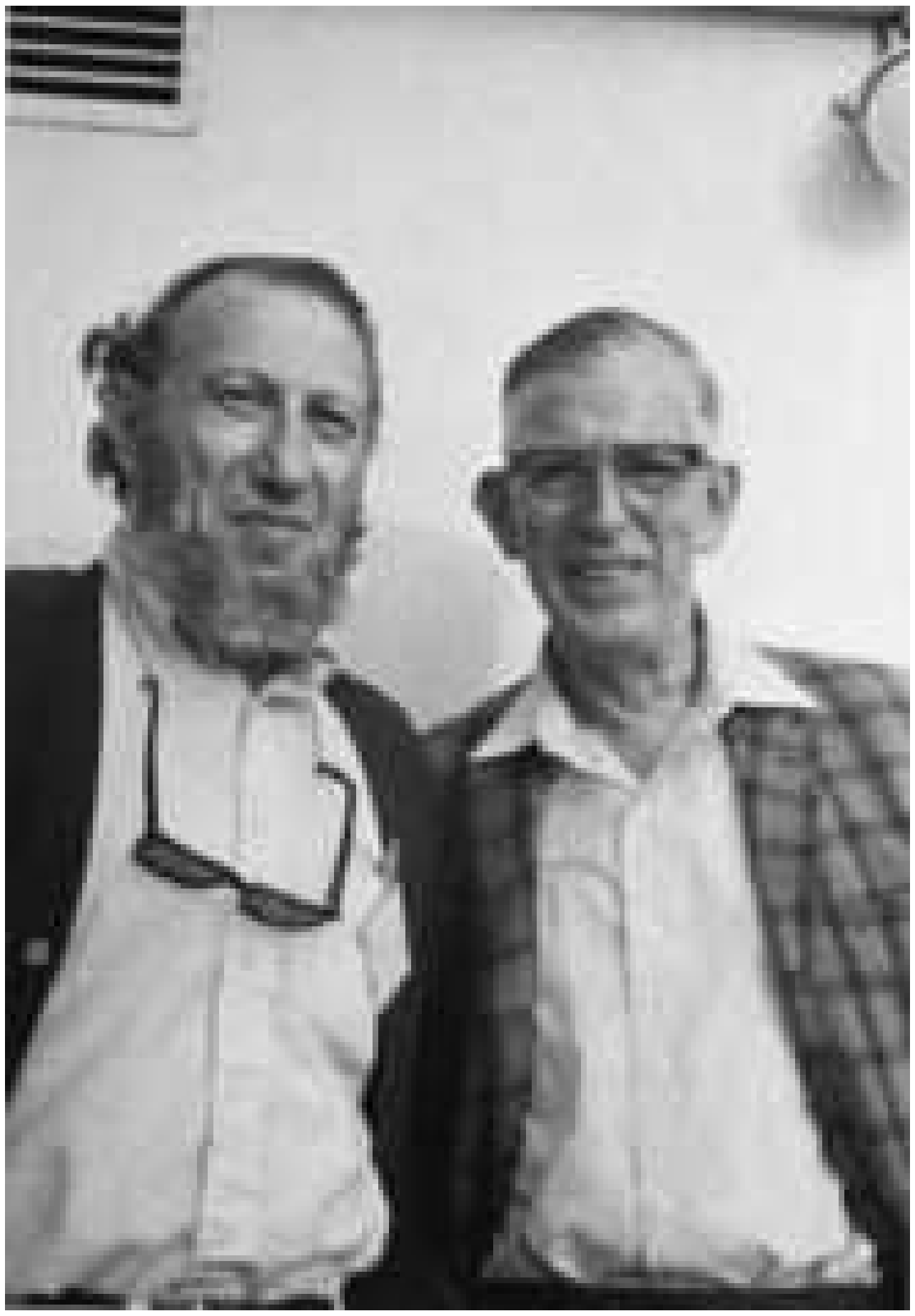

FIG. 4. *Ingram Olkin and Bill Cochran, 1973.*

They were only a few years older than some of the students. We would gather across the street from the department in mid-afternoon. There was a little store where people could get ice cream. In terms of classes, Robbins' batting average was always somewhere in the order of 75% to 80%, by which I mean 75% or 80% of his lectures were really superb. And Hoeffding was a different type of lecturer. Hoeffding's lectures were built-up from the small to the big, whereas Robbins tried to instill some excitement, and to communicate a lot of the key ideas. Hoeffding was much more methodical, as was Bose. They covered the intricacies, and their lectures were not really exciting, but the material was exciting. It was all new. One of the things that people don't quite realize is that there were no books. This was 1948. Cramér (1946) was available. The Kendall books were available (Kendall, 1944, 1946). William Feller's book in probability did not come out until 1950 (Feller, 1950). Robbins had a copy of the manuscript, I suspect, and taught some parts of it, but added a lot. So everything was new. Hoeffding's lectures in nonparametric were all his own which then appeared subsequently as his papers, and the same with Bose and Roy. Hotelling was not a great lecturer. But his lectures were pleasant to hear at times because he was very erudite, and I'd like to say he spoke in prose. There were always complete sentences and they were beautifully put together.

**Sampson:** That's the newspaper man in him.

**Olkin:** That's right. He was a journalist major at the University of Washington, and his writing is clear, and beautiful, and this is true in the papers that he published. Hotelling was an intellectual force. He was interested in pushing students in every possible way and fostering them.

**Sampson:** What do you mean by pushing students?

**Olkin:** Well, let me give you an illustration in my own case. I wanted to take a course in multivariate and it happened that at the time that I wanted to take the course, the course was not given.

I had never taken multivariate. I had no idea why I wanted to take multivariate, but I did. I went to Hotelling and I told him that I wanted to take multivariate, and he said, "Well, why don't you get P. L. Hsu's notes from last year, and study them on your own?" "And then we'll have an oral—you can lecture on it." I've forgotten exactly how he was going to grade it. As it turned out, one of the other students, Walter Deemer, had the same desire. The two of us got Al Bowker's copy of P. L. Hsu's notes, which we went through, and we worked on it. That's where Walter and I recognized that some of the results on Jacobians were things that we could improve and expand on, which we did. But I want to come back to the role Hotelling played. At the end of the quarter, Walter and I gave a lecture on what we had read and accomplished on Jacobians. Hotelling said, "You must publish this." We never thought about publishing this. We were actually very young, and timid, and so we left it more or less in Hotelling's hands. Both Walter and I said this is based on Hsu's lectures, and maybe Hsu should become a coauthor or even author it himself. In any case, Hotelling said he would write to Hsu. At this time the United States did not have international relations with China, and you could not send a letter from the U.S. to China. Hotelling wrote a letter to Egon Pearson and asked Egon Pearson to transmit the letter to Hsu. Later, Hsu replied in the same way and indicated that he did not want to become a coauthor. He thought we had extended the methodology, and that we should publish it. However, he had a few thoughts about one of the theorems at the end, and he asked if we would include several paragraphs that he supplied. We did, and our paper came out in *Biometrika* (Deemer and Olkin, 1951).

In general at every lecture, at every public meeting, Hotelling would publicly mention what students at Chapel Hill had done, so that the profession would know what they had accomplished and in this way he would try to help their careers.

**Sampson:** Did he have social events at his house to encourage interactions among students and faculty?

**Olkin:** Hotelling had an afternoon tea, the second Sunday of every month, which, if I recall correctly, the students labeled "Hotelling's T." Faculty and students would meet at his house. Hotelling had an encyclopedic memory. His conversation was never what you might call small stuff. You would ask him some simple question such as "I see that they're tearing up Franklin Street?" And Hotelling would say, "It was in 1824, that Franklin Street was first developed," and he would give you an entire history of whatever it is you were discussing. Hotelling's wife, Susanna, was a vivacious person who was interested in the students, helped everybody, and was extremely social.

**Sampson:** One has the sense that perhaps Hotelling might be a bit intimidating to have conversations with?

**Olkin:** I don't know if intimidating is right, but you wouldn't talk to Hotelling without recognizing that you were talking to somebody who had achieved a lot. There was a respectful tone that one always maintained with Hotelling. He could generate a respect for the field, and an inspiration for working in the field.

**Sampson:** Did you keep in contact with Hotelling after you left North Carolina?

**Olkin:** Afterwards because of my sense of appreciation for him, when Hotelling turned 65, I suggested to several people from Chapel Hill that we have a festschrift in his honor. Ghurye, Hoeffding, Madow, Mann and I were the editors and we invited many people to submit a paper (Olkin et al., 1960). I recall vividly the response from Joseph Doob. We had asked Doob if he would submit a paper in honor of Hotelling, and Doob said, "Of course I will submit a paper, Hotelling saved my life." As it turned out, in 1939, I believe, Doob was out of a job and Hotelling recognized his talent and gave him a position at Columbia. He gave Doob a job and he brought Wald to Columbia. Not many people would bring someone as good as Wald who might be a competitor. Hotelling was above that in the sense that he was only interested in your scholarly and intellectual affairs. He never entered into any kind of gossip—he was always just basically on an intellectual level.

**Sampson:** Hotelling was your thesis advisor?

**Olkin:** My thesis advisors were Roy and Hotelling.

**Sampson:** What was Hoeffding like?

**Olkin:** Hoeffding was very, very quiet. I'm not sure if he came from Germany or Russia in the 30s, but he was there with his mother. Hoeffding had a sibling. I think he was a journalist who also came over at the time. Hoeffding was focused on the mathematics of the field. He was not a person for small talk, so it was very hard to have a conversation. I mean, you might give a whole sentence, and his reply would be "yes." Then you would have to think of an entire new sentence. You did not get too much response.

**Sampson:** You later had translated from German one of Hoeffding's works that was multivariate in nature.

**Olkin:** It was Hoeffding's 1940 thesis that was on correlations, fixed marginals and bivariate distributions (Hoeffding, 1940). What we call the Frechét bounds are really Hoeffding–Frechét bounds. It's hard to know who did what first, but Hoeffding definitely had the result there.

## RESEARCH AND COLLABORATIONS

**Sampson:** I'd like to now talk about your research. Your published research is quite diverse, but there seems to be one constant. And with very few exceptions, the vast majority of it is published with coauthors. And this began relatively early in your career. It wasn't something that you picked up later. What is it about joint research in statistics that so entices you?

**Olkin:** I think there are probably two factors. One factor is, I think, I enjoy working with people, and you have to enjoy working with people if you're going to coauthor papers. The other is that I found that with coauthors, the sum is really more than what each individual provides. This is because in some way each one complements the other, and so you move ahead much faster. It avoids doldrums. If one person is down, the other person hopefully is up and then you continue producing.

**Sampson:** When I look through your list of coauthors, there is not a whole lot of common threads among them. They are a tremendously diverse group.

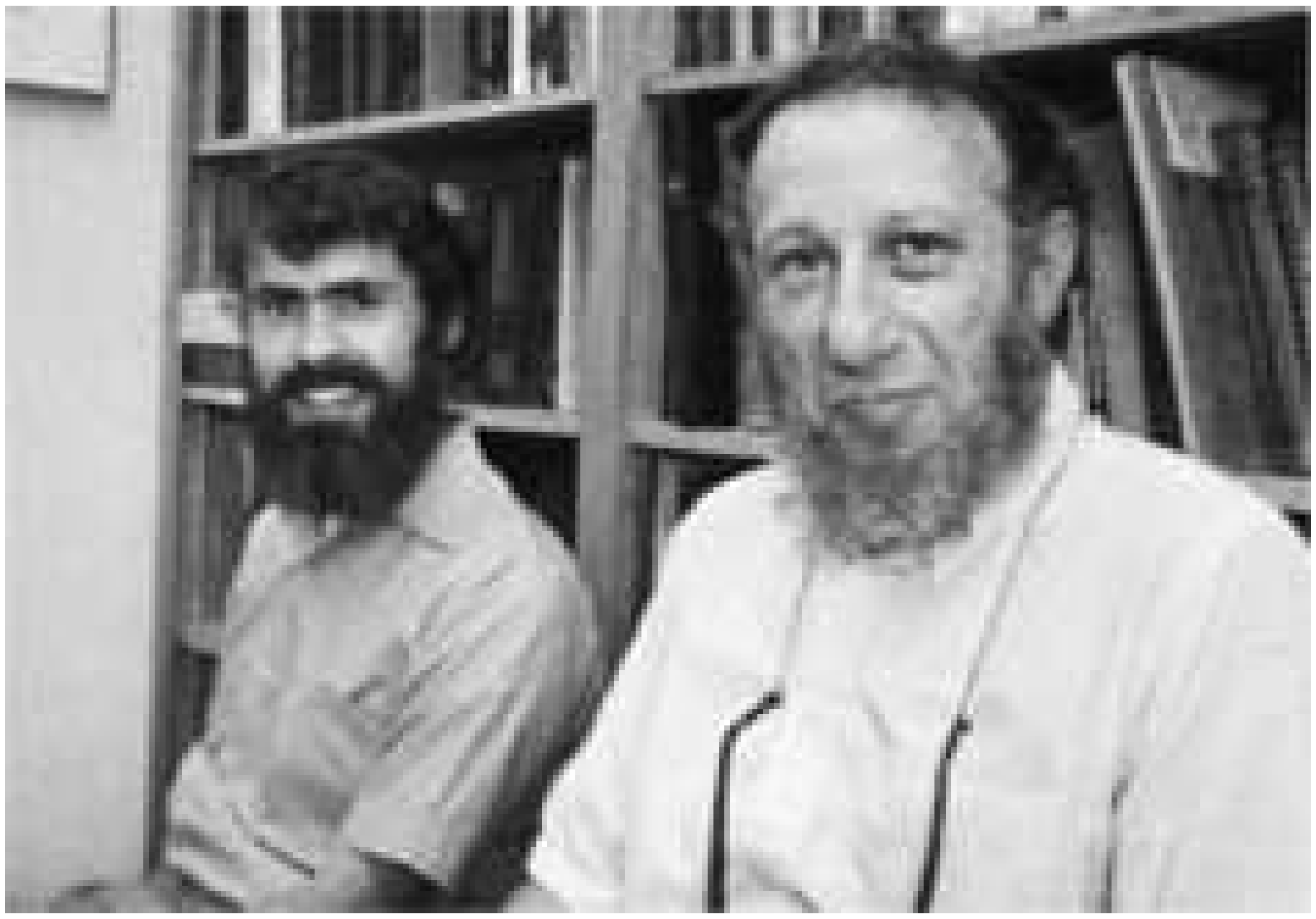

Fig. 5. *Al Marshall and Ingram Olkin, 1973.*

**Olkin:** That's right. Most of the coauthorships were started in an undesigned way. That is, they were not thought of that I would work with someone. Somehow an idea came up. I may have talked to somebody about it, or they talked to me about it. Then suddenly it became really an interesting problem that the two or three of us would work on. There's only one paper that I recall that was done by different groups. There's a paper which Seymour Sherman called the "Chicago Six," and that paper did have contributions from different people without our starting to work together (Das Gupta et al., 1972).

This might be a good point to tell you about a particular collaboration. You mentioned that I had a lot of coauthors, which is correct. But I also have one coauthor with whom I've been involved for 50 years.

**Sampson:** Al Marshall!

**Olkin:** Yes and that happened by accident. I was on sabbatical leave at Stanford from Michigan State in 1958, and I had just published a paper with John Pratt on a multivariate Chebyshev inequality (Pratt and Olkin, 1958). I'm not even sure the paper had appeared. In any case, Al Marshall's thesis, at the University of Washington with Bill Birnbaum was on a multivariate Chebyshev inequality. I believe that I saw an abstract of Al's, and I think I wrote to him about the paper that Pratt and I had published—they did not significantly overlap by the way. In any case, Al, when he graduated, was either a post-doc or a visitor at Stanford, and as it turned out by chance our offices were next door to each other. (As an aside, I still remember the people in the five offices adjacent to me. There was George Forsythe, Bill Madow, myself, Al Marshall, Ben Epstein, and Bob Bechhofer. I could talk to George Forsythe about computing, with Bill Madow on sampling, with Ben Epstein on the exponential distribution, and with Bob Bechhofer on ranking and selection, but mainly with Al.) So Al and I quietly started working together and we wrote 1 or 2 papers, and then I returned to Michigan State. I was back at Stanford in 1961 and we started a collaboration at that time that moved from Chebyshev inequalities, occasionally to other inequalities, and then later to Schur functions and majorization. We were able to continue in a serious way because we both spent a year at Cambridge in 1967–1968. By that time, the seeds of our majorization research had been established. At Cambridge, Al and I gave lectures on majorization and began the work that led to our book (Marshall and Olkin, 1979). From 1967 to when the book appeared in 1979—that's over a 12 or 13 year period—we were collecting results, working together, and then ultimately wrote the book.

**Sampson:** You also spent a lot of time at Boeing.

**Olkin:** Oh yes. I was a consultant at Boeing which means that they made the opportunity available for me to come up whenever Al and I wanted to.

The Boeing Scientific Research Labs was really a great place. It had a galaxy of good people. In addition to Al, Frank Proschan was there, as were Jim Esary, George Marsaglia, Sam Saunders, Roger Wets, Dave Walkup and several others. And the visitors were Dick Barlow, Ron Pyke, Victor Klee, Bill Birnbaum and myself. It was always an exciting place.

**Sampson:** How much time did you spend there?

**Olkin:** My recollection is that we'd try to get together every few weeks. Al would come down to Stanford for a few days, or I would go up to Seattle for a few days. The visits were not extensive visits, but we would often work on the germ of the idea, and then try to develop it a little bit individually, and then get together again.

**Sampson:** And that was before email back and teleconference made communications so much easier.

**Olkin:** Everything was handwritten, of course. However, you do have to recognize that it's very difficult to write a book just by going back and forth this way. So we actually spent time together away. Beside the year in Cambridge, we were also in England for a year and we were in Augsburg and Zurich for three months together. We were in a number of places for more extended periods then just these visits.

**Sampson:** Did you spend time in British Columbia?

**Olkin:** I spent a quarter there, when Al was on the faculty at the University of British Columbia, Also I did visit on occasion for shorter periods. What we tried to do was work in short spurts and then have culmination by having longer periods together.

**Sampson:** You also did with Al a fair number of papers on distributions, and some more recent work has been on families of distribution. And of course there's the well-known Marshall–Olkin bivariate and multivariate exponential distributions (Marshall and Olkin, 1967) that are widely used.

**Olkin:** We started working on the question of how to generate bivariate distributions that have certain kinds of nice properties, and our first instance of that was what's now called the Marshall–Olkin bivariate exponential. More recently we planned to try to write a book on nonnormal bivariate distributions. We started, and then we said we'd better first write a chapter on univariate nonnormal distributions. Well we started that and we found that we wrote one chapter, and then another chapter, and then another chapter, and we found we never got to multivariate distributions! We are now about to publish a book entitled *Life Distributions: Nonparametric, Semiparametirc and Parametric Families.* The book is almost complete. We're in the process of preparing the graphs for the book, and when that happens, the book will be finished. So my guess is within the next three months, we will be sending it off for publication.

**Sampson:** To go back to your work on the exponential, my recollection is that before yours and Al's work, people did not do a lot of work looking at properties of the exponential distribution in the univariate case and trying to extend them. Was yours the first that looked at the memoryless property and tried to extend it to two-dimensions?

**Olkin:** I think in that respect, ours was the first. However, I think your point is really an important one. What we tried to do is look at a property of a univariate distribution and see in what way one could extend that particular property. Not all univariate properties have nice extensions to the bivariate case. The memoryless property was one that had a nice extension. But there are lots of other examples where we extended the ideas of the characteristics of the univariate distribution to the bivariate.

**Sampson:** More generally, you have a real interest in solving functional equations.

**Olkin:** Functional equations have stood us very well. I'm really a proponent of Aczél's (1966) book on functional equations. Al and I just submitted a paper in honor of S. N. Roy that's coming out in a special issue of the *Journal of Statistical Planning and Inference.* And there we solve a number of functional equations. You might ask the question, why is the Weibull distribution so popular and how did Weibull come about it? Well, it turns out that Weibull actually came to it from data analysis. However, there is a nice rationale in terms of the distribution as the solution of a functional equation. When you have scale-parameter families and proportional hazard families, the Weibull is the coincidence of the two families. This uses a functional equation solution again.

**Sampson:** Ingram, what do you see as the secret of your collaboration with Al in terms of working style? I think you told me he is a craftsman by nature and for instance loves to build furniture. Is that style something that carried over into how you two interact?

**Olkin:** Well, Al is definitely more of a mathematician than I am, but I don't think that's the issue.

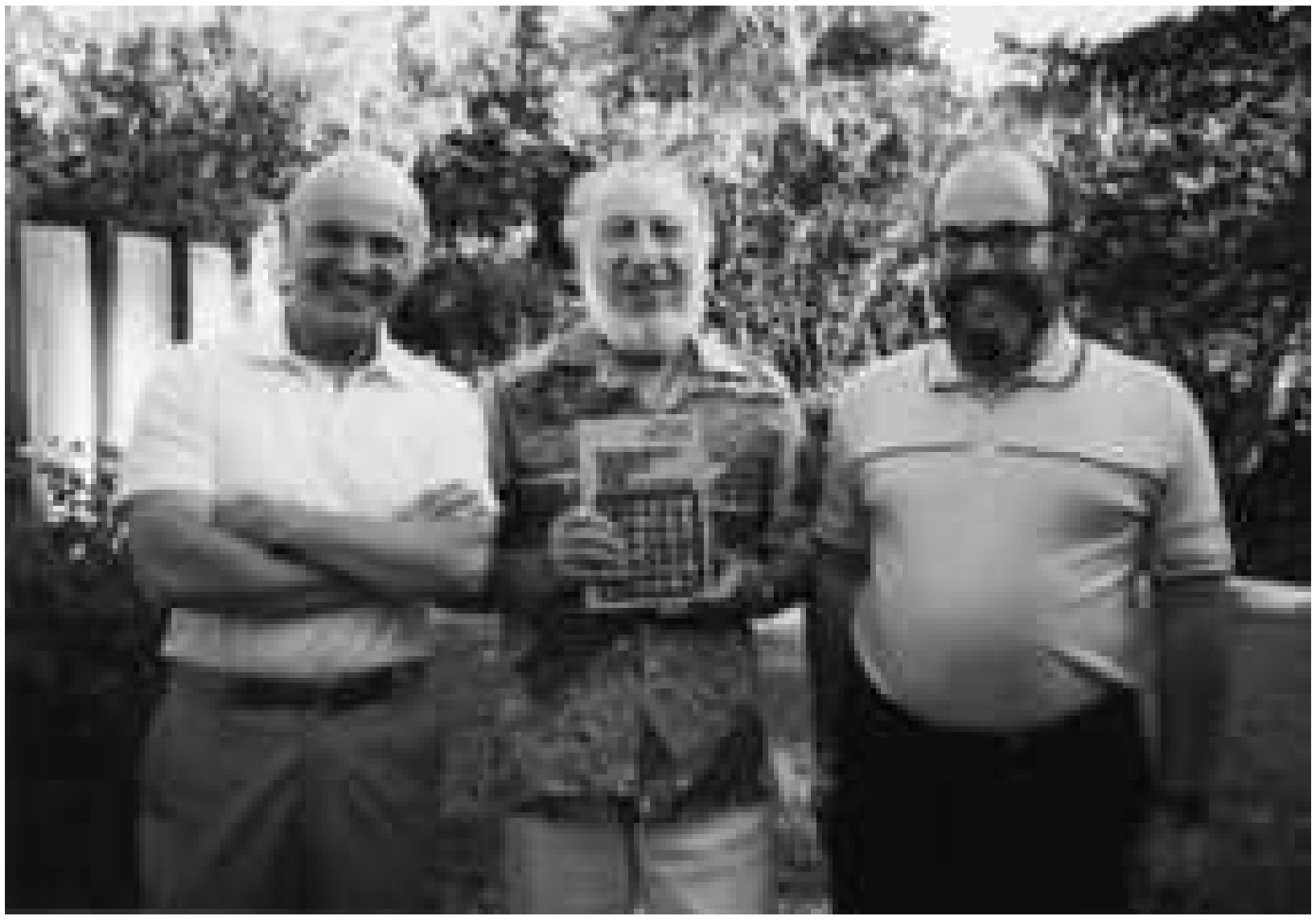

Fig. 6. *Cy Derman, Ingram Olkin and Leon Gleser, 1986.*

I think that Al has a unique way of looking at problems, and it happens to be slightly different from mine. And so Al brings an expertise that I would not have, but I bring a different view of some of the problems that Al wouldn't have. I think it's fair to say that Al probably could write the book without me, and that I probably could not write the book without him. But that the end result is very different from what it would have been had he just written the book alone. I think the wide applicability is something that comes about from our collaboration. If you think about the majorization book, there are a lot of technical aspects, but there's a lot of breadth and scope—applications in probability and statistics, in matrix theory and in combinatorics. I think a lot of that is the kind of thing that I would bring to the book, and a lot of the results would be things that Al developed on his own or we did together.

**Sampson:** My sense is, in addition, that the two of you were good personal friends. You share a lot interests in common.

**Olkin:** We do. We go to concerts together, and our wives have gone together on different trips, and we've gone together on different trips. We used to do this a lot more than now because we each have large families that we're involved with. We haven't visited each other as much as we used to, but in the early days we certainly spent a lot of time together. Our collaboration has been unique in my life in that it has extended over such a long period of time and over so many different papers and two books. It's been very, very fruitful.

Indeed, some people think we're one person whose name is hyphenated as Marshall–Olkin. I have to tell you that Al once sent me a CD that he had found. The cover featured music by a composer whose name is Ingram Marshall.

**Sampson:** You've had a number of other coauthors that you've done a lot of papers with including Leon Gleser and Larry Hedges, both of whom were your Ph.D. students.

**Olkin:** Leon and I overlap a lot in interests and skills, and that's a very nice collaboration because we're both attuned to the same kind of orientation. Leon is really a very good problem solver and together we've worked on quite a number of different problems. I've always enjoyed that collaboration. I've probably published more with Leon, after Al, than with anybody else.

## META-ANALYSIS

**Olkin:** Larry and I started working together as a result of my appointment in the School of Education at Stanford and my early involvement with meta-analysis. During my life I've tried every ten, fifteen years, to become involved in something a little different, a little orthogonal from what I had being doing. Because once you continue it's very hard to keep up the excitement in the field after you've been publishing, and it's nice to start on something new. And meta-analysis was one of the areas that was very different. And that started in a very innocent way. It began because a colleague in the School of Education said to me, "Ingram, there have been literally hundreds of papers on the effects of certain teaching aspects, but these studies are small and they're not significant. Is there any way to put together this mass of individual studies?" So that started me on the field of meta-analysis. That was in the early 70s, and I wrote one paper in the 70s by myself, and then I started with students. And Larry was the first student. He wrote a dissertation on meta-analysis.

**Sampson:** He was a student in Education?

**Olkin:** Yes, at that time we had a program in education. It was comparable to what you might call biostatistics, only it was educational statistics in which the students would take all the courses in the Statistics Department, more or less through what would be a normal Ph.D. program, but the dissertation would be on an educational topic. They may not have taken every single course, but they took most of the courses. Larry was in that situation. And we started writing this book on statistical methods for meta-analysis.

**Sampson:** Was that while he was still a student?

**Olkin:** No. The book was published in 1985 (Hedges and Olkin, 1985), and my recollection is he may have finished his Ph.D. in 1981. But we started working on it almost immediately thereafter. And the reason we started working on it was we read many of the papers that were being published, and we recognized that there was not much statistical methodology being used in them. There was a book earlier than ours by Glass, McGraw and Smith (1981), and it had some statistics. But it didn't have a systematic statistical development. So that's what we provided in our book. Our book seems to have had a catalytic effect on meta-analysis. Afterwards my increased involvement with meta-analysis brought me into the sphere of the medical profession. These were really the people who are doing medical research and trying to come up with conclusions about the state of their field, so it has been very exciting for me.

**Sampson:** You've given and still continue to give short courses on meta-analysis.

**Olkin:** The short courses really started with one of the medical originators of meta-analysis, a physician by the name of Tom Chalmers. Chalmers was a delightful person, and he and I met when he was at Mt. Sinai Hospital. He was the head of what I think was called a Technical Assessment Division. And they had a site visit, and I was on that site visit committee. I enunciated a number of factors in favor of meta-analysis that the hospital, at the time, hadn't recognized. As a result of that, Chalmers and I became friends, and we started giving some short courses. It was a marvelous collaboration because he would speak for an hour and a half on medicine and then I would speak for an hour and a half on the statistical aspects of the studies. And then we would go back and forth, and we did this for two days, at several different places.

Chalmers started his lecture in a way that I couldn't because I'm not a physician. His first few sentences were something like, "I got into the field of meta-analysis because I realized that I was killing patients." Well, of course, when a physician says that it creates quite a stir in the audience. Tom had a very wonderful way of presenting the material. Unfortunately, he died a few years ago of prostrate cancer. One of his protégées is a physician that I work with a lot now, who's very good, and a lovely colleague. That's Joseph Lau, who's at the New England Medical Center. Joseph has a group working there, and they're one of the producers of a lot of meta-analyses.

**Sampson:** Where have you given some of your short courses—particularly the international ones?

**Olkin:** To start with, several were for the American Statistical Association. I've given short courses in Singapore and in Hong Kong and I've given short courses throughout Europe, including Switzerland, Spain, Croatia, Holland, and Austria. I may have left out a few places—I have not given a short course in France. Some of these courses I taught with Joseph Lau.

## EDITORIAL CONTRIBUTIONS

**Sampson:** Ingram, let's talk about your editorial work. You've had a life-long involvement in various editorial capacities, in statistics journals, educational journals, and mathematics journals. Why have you chosen to devote so much energy to these purposes?

**Olkin:** Well, earlier on I mentioned my mother took me to the library.

**Sampson:** You are blaming it on your mother?!

[**Laughter**]

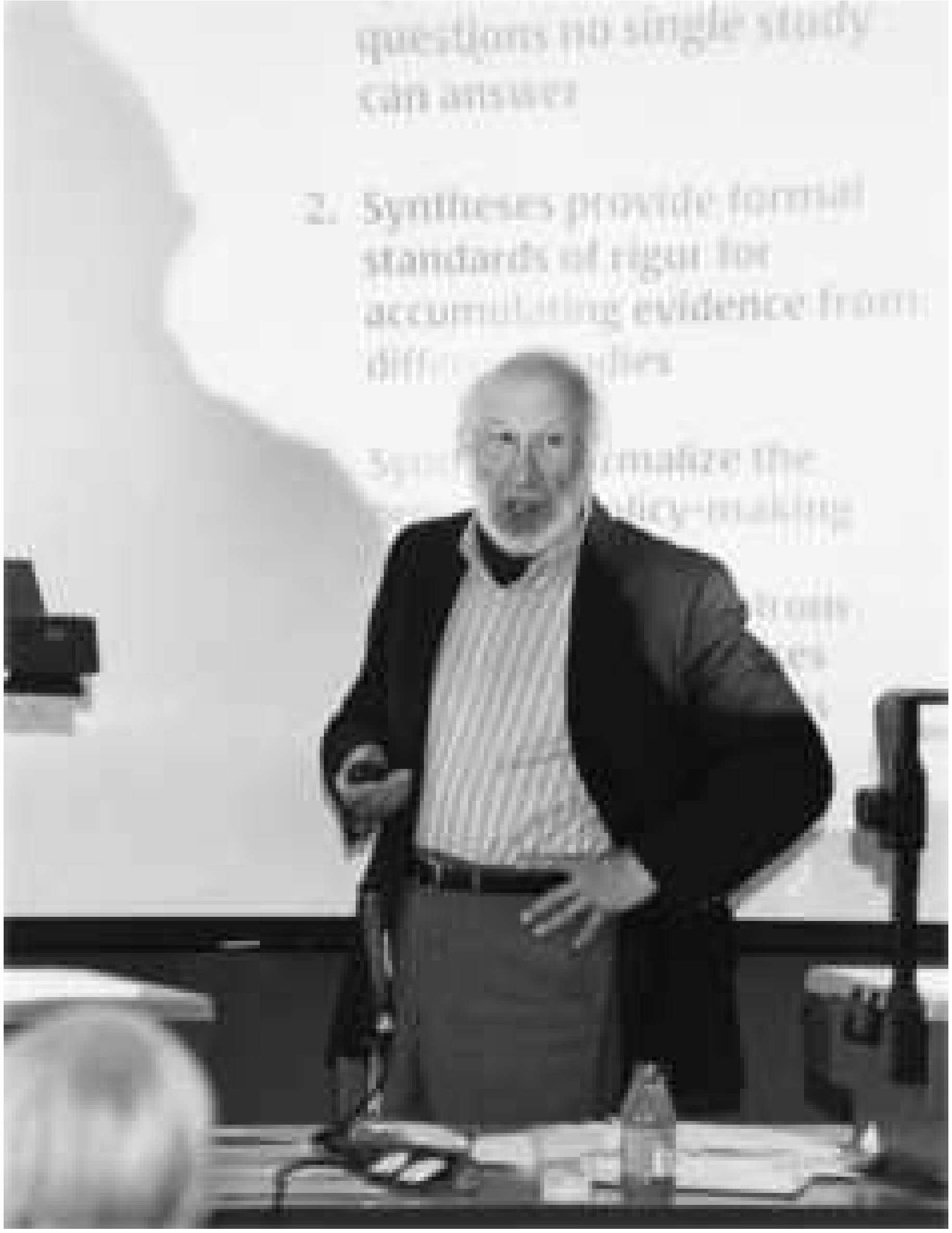

Fig. 7. *Ingram Olkin speaking on meta-analysis, University of Pittsburgh, March, 1998.*

**Olkin:** No, I'm not going to blame it on my mother. I don't know why I was so involved in journals. When I was at Chapel Hill in 1946, it was not a major enterprise to read all the journals. They would come to the library and I would browse through the journals. I found that a lot of times browsing through an article would ring a bell about something that was similar to what I was doing. That was the beginning of my interest in journals. At Stanford, I was on the Library Committee and I was always involved in trying to build up the library. But that doesn't relate to the editorial work. I was an associate editor for *JASA*, but that also wasn't the real catalyst. It was when I became editor of the *Annals* that I started thinking a lot more about journals. What became clear to me is that the growth in statistics from the time I graduated in 1951 to the time I was an editor of the *Annals* in 1972, had resulted in the *Annals*' publishing over 2,000 pages a year. No editor could really review all the papers. The editor was really a manager. You just shuffled the papers to different associate editors and you intervened in questionable cases. Otherwise, when things were clear-cut, you merely accepted what the associate editor suggested. Probability and statistics were both growing and it was just unreasonable to think of trying to keep the two together. Now, there was a controversy at this time. The question was should we begin to splinter into several separate groups? There were camps on both sides. There were clearly people who thought that the *Annals of Mathematical Statistics* should include both statistics and probability. I thought from a practical point of view, the growth was too big in both fields, so I proposed splitting the Journal. We went to the Council. After discussion, we did split the Journal. I remained as editor of the *Annals of Statistics* and Ron Pyke became the first editor of the *Annals of Probability*. Of course, in the year 2005, no one would ever question going back to a single journal. The two fields have grown and, in fact, they've splintered even more into the *Annals of Applied Probability* and now into the *Annals of Applied Statistics*.

**Sampson:** I'm curious, was opposition to the split more from the statistical types or the probabilists?

**Olkin:** The statisticians did not want probability split off. But more than that, they also felt that splintering of the field was not good. I was not of that opinion for a variety of reasons. What I saw was that when the society decided not to publish a topic journal such as the *Journal of Multivariate Analysis*, or the *Journal of Time Series*, or the *Journal of Sequential Analysis*, the commercial publishers went into that field. And I felt it was better for the society to sponsor these topic journals, but others felt that this splintered the field. My position was that it was going to be splintered by the commercial publishers in any case, and that I would rather it be splintered within the Society's control. I lost that battle.

Also the beginnings of splintering led me to start to think about journals that the Society might feel comfortable with. Let me go back in time for a moment. There was a journal in engineering, *Techometrics*; there were several journals in biology, *Biometrics*, and to some degree *Biometrika*; there were journals in psychology, *Psychometrika*; and journals in economics, *Econometrica*. There was nothing in education, so Mel Novick who was in psychometrics at Stanford and I talked a lot about possibly starting a journal in education. He and I agreed on two principles; one was that it should be published by a society, even though commercial publishers had approached us; and the second was that we agreed that it would be better for this to be a joint venture between education and statistics, rather than either one taking it on separately. Mel's task was to go to the council of the American Educational Research Association; my task was to go to the board of the American Statistical Association and to see if we could propose a joint venture. That was achieved and Mel Novick was the first editor—I had not wanted to be involved as an editor—and the *Journal of Educational Statistics* (now called the J*ournal of Educational and Behavioral Statistics*) was launched in 1976.

**Sampson:** What were the issues involved in the creation of *Statistical Science* and how did it get started?

**Olkin:** That was very interesting in a variety of ways. A group of statisticians, without any authority from societies, would get together. The group included National Science program directors, because they were on top of the problems in the field, and also people who were interested in doing something for the society. I hate to mention members because I'm going to leave out some, but I'll mention Jerry Sacks, Morrie DeGroot, Steve Fienberg, Peter Bickel, Paul Shaman and myself, and people such as Bruce Trumbull and Nancy Flournoy. Both Bruce and Nancy were NSF program directors as had been Jerry Sacks and Paul Shaman. One theme that kept surfacing was the need for some kind of generalist journal. Morrie and I, in particular, were the leads in that. Morrie was interested in being editor, and that was great because he was a really a superb editor. We both had in mind doing some history, doing some interviews, and doing some general papers. What we firmly believed was that a paper that was suitable for *JASA* or the *Annals* would not be suitable for *Statistical Science*. We wanted these to be generalist type articles with a wider readership

**Sampson:** Why did it end being published by the IMS?

**Olkin:** I think that was a decision that we made. It was the practicality of the situation. We could get it through the IMS and we could not get it through the ASA. It would involve a different type of proposal to get it through the ASA. We were all members of IMS. A lot of us were on the council; one of us was a treasurer. So we could get a positive reaction from the Council of the IMS, and as it turned out we did.

**Sampson:** *Statistical Science* is almost twenty years old. Has it successfully fulfilled what was the original ambition for it?

**Olkin:** When I talk to people, they say they love the interviews. They read other articles on occasion.

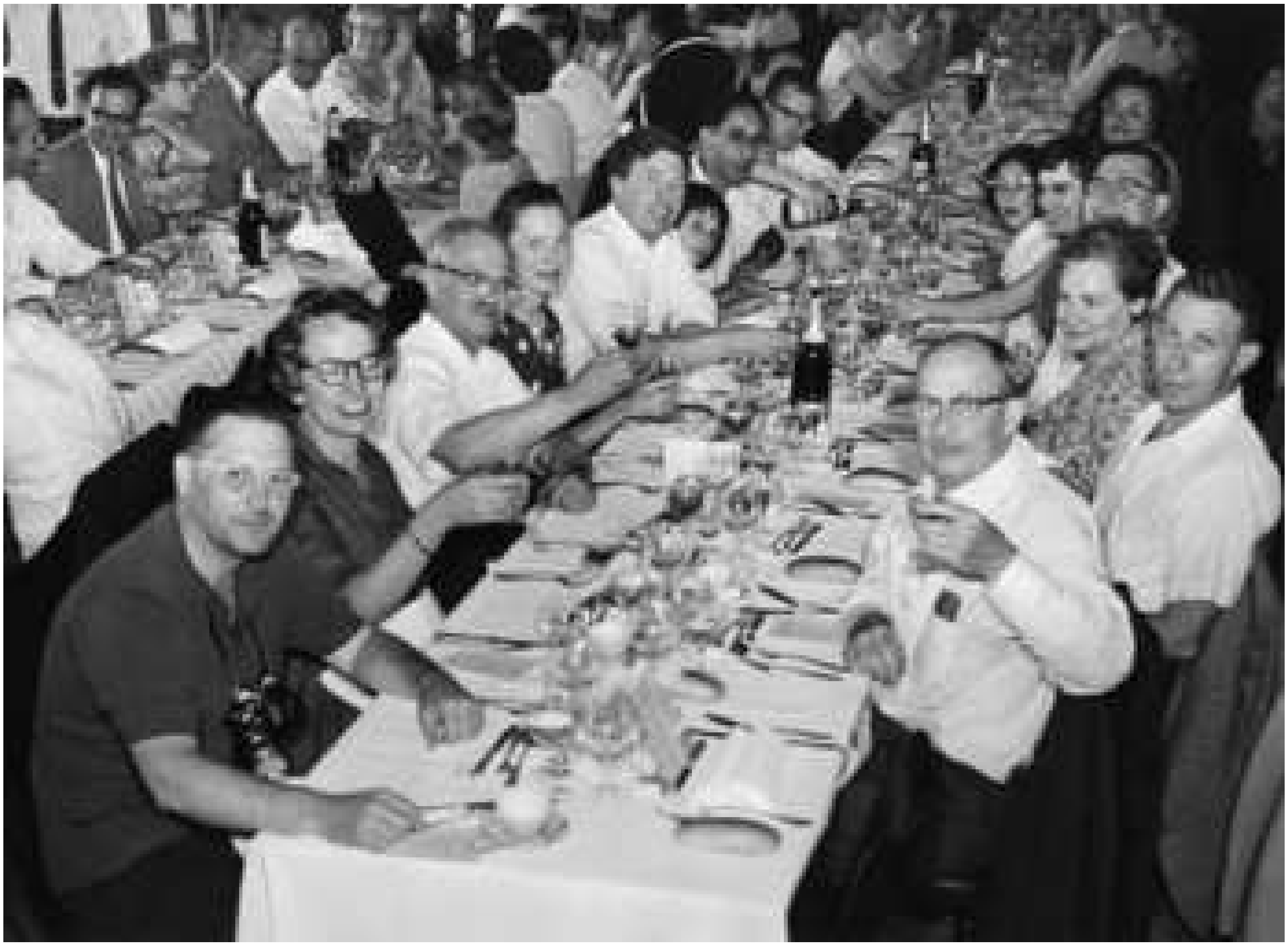

FIG. 8. *ISI Meeting, Paris, 1961. Left side: Unknown, Elizabeth Scott, Jerzy Neyman, Anne Durbin, Jim Durbin, Miriam Chernoff, Herman Chernoff, Unknown; Right side: Jack Youden, Ingram Oklin, Dorothy Gilford, Manny Parzen, Carol Parzen, Ellen Chernoff, Judy Chernoff.*

It's not a journal that people read from cover-to-cover. But almost everybody likes something, and I have found very few people who say that they dislike the journal. A testament to its success, I think, is that a lot of members of the American Statistical Association are sorry that it's not a journal that's joint.

**Sampson:** Do you think *Statistical Science* has kept the mathematical content far enough away from *JASA* and the *Annals*?

**Olkin:** I think that it is difficult in a lot of subjects to not become too technical. But I think, by and large, that there are enough articles that are not mathematical, so that people continue to enjoy reading it, and keep referring to it.

**Sampson:** These days, one has a sense that there's going to be more of a movement to electronic journals, e-journals, that may never exist on paper. Do you have any thoughts on this?

**Olkin:** Well, it's clear that the professions have to face the issue of electronic journals. One of the questions, of course, for statistics, is that statistics is an archival science. Not every field is archival. We want to be able to preserve what the journals produce. I think one of the problems that will have to be faced is who is going to preserve them and in what condition will they be preserved? But certainly electronic journals will continue to become the natural format.

**Sampson:** You mentioned earlier that the for-profit publishers are putting out the journals in more focused areas of statistics. Would you be advocating that the societies today try to pick up that role, too?

**Olkin:** I have tried to get societies to focus a little bit on some of these splinter topics, but by-and-large societies are conservative. I think they fear that if they publish splinter journals, the number of such splinters could be large. And they can't quite face why, for example, the IMS should publish a journal in design of experiments, and not on multivariate, and not on sequential, etc. I think they are faced with the plethora of possible splintered topics. It was perhaps somewhat tolerable when different publishers picked up different parts of the splintering, but as time has evolved, these separate companies have merged. So now, all of a sudden, instead of, for instance, ten journals from ten publishers, we have ten journals from one publisher—a super publisher. All of these splintered journals are now creating a really serious financial problem for any library.

**Sampson:** These super publishers have very strong price leverage.

**Olkin:** That's, in effect, what is happening. And we did not foresee the mergers. What we saw were the splinters. But we did not see the fact that these would all coalesce under one or two key publishers. The problem is already very acute in medicine because the splinters have further specialized. There might be a highly specialized journal, and if there is a single doctor studying that particular area at an institution, their library may be forced to purchase a bundle of journals by that publisher in order to obtain the one specialized journal.

## STATISTICS ON THE NATIONAL SCENE: NAEP, NISS AND NIST

**Sampson:** Let's talk about your work in statistics and its influence on national policy issues—in particular, about your work with the NCES (National Center for Education Statistics), NAEP (National Assessment of Educational Progress), NISS (National Institute of Statistical Science) and NIST (National Institute of Standards and Technology).

**Olkin:** I would like to talk first about the National Center for Education Statistics. There are a number of governmental agencies charged with the collection of data relevant to their particular area. There's the National Center for Education Statistics, National Center for Health Statistics, the Bureau of Labor Statistics, Agriculture, and so forth. One of the things that became clear early on was that even though the National Center for Education Statistics has the word "statistics" in its title, the number of whom we would call professional statisticians in NCES is very small. Some of the other agencies have many more statisticians for a variety of historical reasons.

In any case, in the late 1980s I had contact with Emerson Elliot, the Commissioner of the National Center for Education Statistics. We talked a lot, and I commented to him on the fact that not that many statisticians moved in the sphere of education, and I thought that it would be good if that could be changed. At that time, the American Statistical Association already had a fellowship program—I believe it was in existence with the Bureau of Labor Statistics. In any case, I suggested to Emerson that there be a fellowship program at NCES. And Emerson was a very positive person in trying to do all kinds of good things for NCES, and he thought he would be willing to put up some financial support. The ultimate conclusion to our discussion was that a fellowship program was started. He asked me if I would be willing to help start it, and I said I would. I was instrumental in getting Larry Hedges and Ed Haertel to be Fellows during the early period. Subsequently, Julie Shaffer was a Fellow, Jeremy Finn was a Fellow, and there have been many others. And I think this brought a bit more connection between statistics and education.

But it was also at NCES that I became involved in some of the technical issues in the National Assessment of Educational Progress (NAEP). NAEP has been in existence for many years, and is really one of the fundamental barometers of the state of education in the United States. Lyle Jones and I recently edited a history of NAEP, often called the "Nation's Report Card" (Jones and Olkin, 2004). It is a fascinating story. The story starts in the early 1960s with Francis Keppel, then U.S. Commissioner of Education, who recognized the need for a national assessment of education. Keppel was a friend of Ralph Tyler, then director of the Center for Advanced Study at Stanford, and of John Gardner, then president of the Carnegie Corporation of New York. The three had talked about the idea of a national assessment. Keppel asked Tyler to suggest a way to evaluate education, and Tyler convened a committee consisting of John Tukey (Chair), Robert Abelson, Lee Cronbach and Lyle Jones to develop a plan for a periodic national assessment. Gardner, via the Carnegie Corporation provided funding for two conferences. From this beginning, through a series of conferences, committees, and partial assessments, an assessment of 17-year-olds in citizenship, science and writing took place in 1969. We have come a long way from that point, and it is a credit to NCES that NAEP has maintained credibility throughout its history. Today we have in addition to the national NAEP, a state NAEP, because states are interested in how well students in the state are doing. What all of this shows is that it takes the confluence of many forces to accomplish a program of this magnitude.

As a result of my education contacts, I became a member of what was then called the Technical Advisory Committee to NAEP, and is now called the Design and Analysis Committee of NAEP. I've been on that Committee for probably 15 or 20 years. It's had somewhere between 10 and 14 members, mostly statisticians and psychometricians—all very good people. We meet three times a year, and our task is to help in some of the technical intricacies in doing a national assessment. I'll give you an example of one of the technical problems.

When Congress mandates that a change be made, this affects the way the testing will take place. For example, Congress mandated that testing give accommodation to children who have disabilities. This meant that NAEP must decide how to deal with children who have hearing problems, or eyesight problems, or dyslexia, or a variety of other problems. But the point is that now if you're doing a trend and suddenly at a certain point Congress makes a change, how do you maintain the trend given this change? We've had many discussions on that. Another example of a technical problem which is of interest is the state NAEP. We have fifty states, and now if you're going to make comparisons between states, we might have more than 1,200 comparisons. So, the question is how should we make these multiple comparisons?

The real problem is how to reconcile a technical correctness with an interpretive correctness. If you do multiple comparisons you might find that state A and state B are not statistically different. But if you don't do multiple comparisons, you might find out there's a significant difference. The issue is now that a legislator in some state will say, "How can there be two answers—we're different or we're not different?" Well, as you know, the multiplicity of acceptable statistical analyses is a standard problem for statisticians in many contexts. It's not just basically a commentary about multiple comparisons versus none. Somebody does a $t$-test and somebody does a nonparametric test, and you could get different answers. We have a problem in interpretation, and the National Center of Educational Statistics has the task of telling the nation its results. So, if you look at the reports, you'll find a variety of suggestions that have been implemented to try to be clear to the public as to what's going on. We've used footnotes to try to explain, and we have continuing discussions on this point.

**Sampson:** NISS is another statistical enterprise that's focused on issues in public policy.

**Olkin:** I'd like to go back to the history of how NISS actually became NISS. The group involved with starting *Statistical Science* also recognized that there was a deficiency in cross-disciplinary research. Ultimately, there was a proposal that Jerry Sacks and I submitted through the IMS to the National Science Foundation to have a panel to discuss and study cross-disciplinary research. When that panel's report was completed, a little booklet about its findings was issued (Olkin and Sacks, 1988). That booklet has been used by many chairs of departments when talking to deans. It's been used in a variety of other contexts. In any case, there was a long discussion in the booklet about the need for cross-disciplinary research. A confluence of many events are needed for an organization like NISS to be formed. The Panel's report provided a rationale for an Institute of Statistics, which, I think, it is fair to say, was my idea. Jerry Sacks helped with the planning and conceptualization and Nancy Flournoy, the NSF Statistics Program Director, managed to provide money for a feasibility study, which resulted in a call for proposals. All in all, it took several years before NISS came to fruition.

**Sampson:** There was a heated competition for NISS. Pittsburgh was one of the finalists, as you know, along with North Carolina. Our state legislature wasn't as generous as the State of North Carolina. The North Carolina group did a wonderful job in obtaining NISS.

**Olkin:** That's absolutely correct. There were five proposals of which two were finalists. One was from a consortium of the University of Pittsburgh and Carnegie Mellon University. Dick Cyert, then President of Carnegie Mellon, was instrumental in offering us space in the future. Carnegie Mellon was building up. And the other proposal was from a consortium in North Carolina, namely, of Duke, North Carolina at Chapel Hill, North Carolina State and the Research Triangle.

At that time, the Research Triangle offered land in their park, but the most critical point was that the North Carolina consortium was instrumental in getting their state legislature to offer quite a bit of money. There was a lot of start-up which was difficult to come by, and so NISS ended up in North Carolina. But in contrast to some of the mathematics institutes, NISS was not designed to be an institute focused on a specific area, but rather on cross disciplinary areas that would have a policy impact. In fact, in the original call for proposals, we limited entries to east of Chicago because we thought that if NISS's work was going to involve policy that it should be within access of Washington, D.C. NISS was started and now is in its tenth year. And I think it has been very successful.

**Sampson:** Also now it's involved with SAMSI which is headed by Jim Berger, and they are housed in the same building.

**Olkin:** I think it's very good for statistics. I would only argue that it's a shame that we don't have three such institutes because I think the profession is growing to such a degree that there should be. A very natural marriage would be between statistics, genetics, and biology to have a more focused role in thinking about how statistics can participate in what is clearly going to be an important and growing field. If statistics isn't involved in the early stages, it may not be involved later on.

**Sampson:** Let's move from NISS to NIST. Briefly, what were your interactions with NIST?

**Olkin:** Well, I was involved with Churchill Eisenhart, who was the first leader of the Statistical Engineering Group at the National Bureau of Standards. Now you have to remember that the National Bureau of Standards had a long history of statistics. In the 1950s, it had on its staff people such as Marvin Zelen, Frank Proschan, Richard Savage and Joan Rosenblatt.

After Churchill Eisenhart retired, Joan Rosenblatt was the head of this thriving group. I was invited to participate in some of their activities. I actually gave a lecture on meta-analysis there, and over the years I've been involved in number of projects. The most recent one I've been involved with is an update of what I'll call the Abramowitz and Stegun (1972) book, *Mathematical Functions*, which is one of the famous and best selling books ever in mathematics. In it there's a chapter on probability and statistics by Marvin Zelen and Norman Severo. This chapter is being updated and I've been involved with a number of people on that project.

## STANFORD

**Sampson:** Ingram, we haven't yet talked much about your university careers. I know you started at Michigan State and then went to Minnesota before Stanford. Also my rough calculation is that Stanford's Statistics Department, is about 60 years old. You've been there for approximately 45 years, which is 75% of the departmental life history and this might give you an interesting perspective.

**Olkin:** Let me talk a little bit about each piece, if I may. The first point is in 1951, when I graduated, there were very few statistics departments. Almost everyone who graduated, generally would go to a math department. What I didn't recognize at that time was that many of us who went to math departments ultimately tried to generate statistics departments. So this started this general growth. When I went to Michigan State, there were not many jobs and I did not have many offers. Math departments would want one statistician, and so if they had one, they weren't ready for a second one. And a lot of math departments didn't have any, and didn't want any. In any case, Leo Katz—once again random chance mechanism plays in your life—was a visitor at Chapel Hill. We became friendly and he said "Well, when you graduate send me your vita." I did, and they offered me a job. Leo Katz was the first member of the statistics group at Michigan State. I cannot remember whether I was second or third, but the other one was Kenneth Arnold who had been in Wisconsin. After that was Jim Hannan. Leo was very good as a manager, and he arranged to get a lot of visitors, such as Herman Rubin and Esther Seiden (who later was on the faculty). R. A. Fisher was a visitor during my stay there. I'm one of the few members of the profession now who can say that R. A. Fisher had dinner at our house.

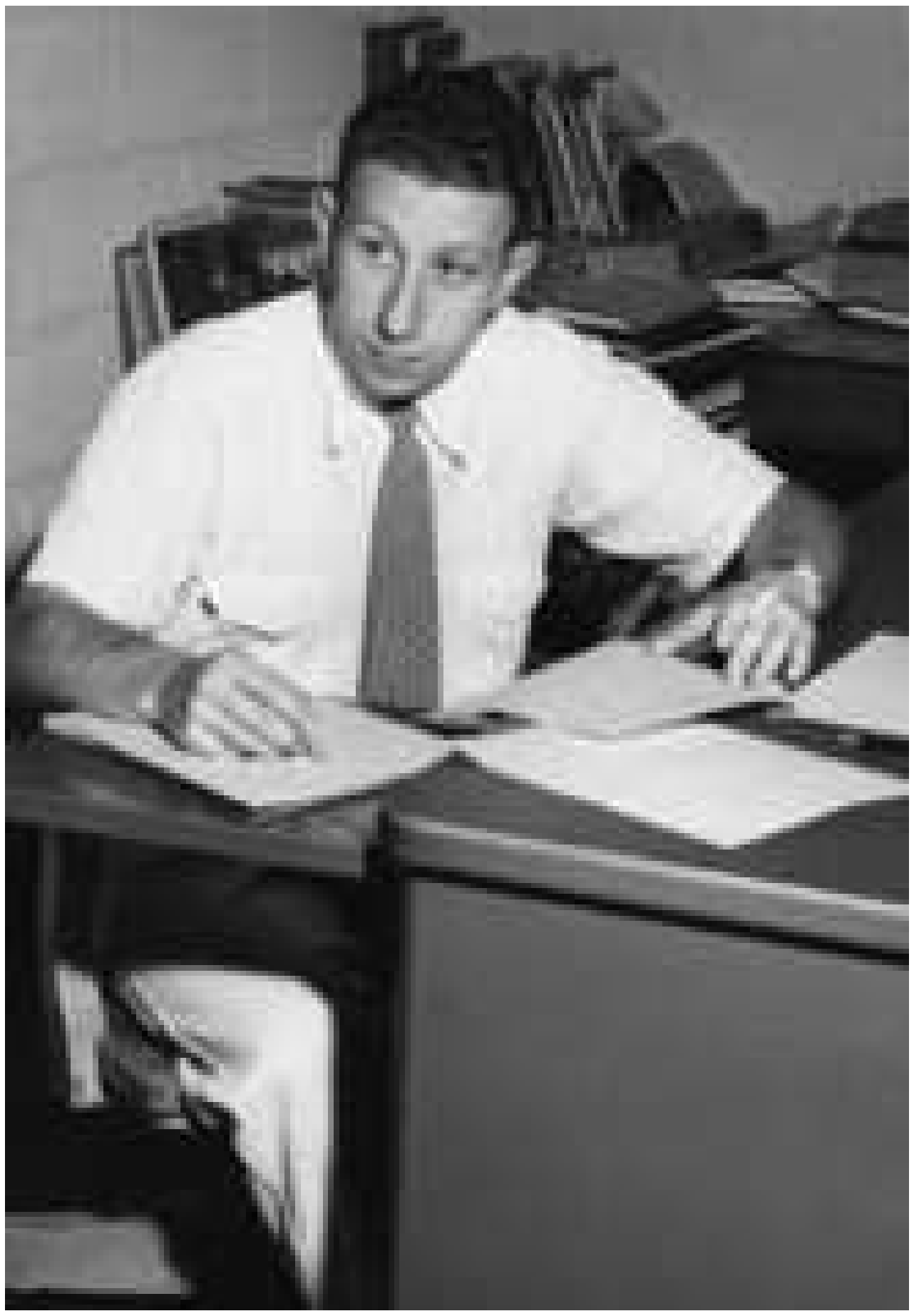

Fig. 9. *Ingram Olkin in his Michigan State University Office, 1950s.*

**Sampson:** That sounds like a story for another interview!

**Olkin:** You're right. In any case, once we had a core group, we were able to form the department, and that was probably by 1956 or 1957. And I really enjoyed Michigan State very much. It was a great part of my life. The state of Michigan had three statistics groups. There was the Michigan State group. Don Darling, Paul Dwyer and Cecil Craig were at Michigan. Milton Sobel and Ben Epstein were at Wayne. We had joint meetings and seminars on occasion where we rotated the places. It was a very good time for statistics. In 1960, I was offered a very nice position at Minnesota, and I decided to accept that. As it turned out, the chair at Minnesota was Palmer Johnson who was an educational statistician—a wonderful person. He died shortly after I arrived and I became the chair. Richard Savage was there, as were Bernie Lindgren and Leo Hurwitz, and we were starting to build up. We hired Milton Sobel and we hired Meyer Dwass. (Meyer later decided to go back to Northwestern.) So this started an increase in Statistics. Later Kallianpur was hired and a number of others. And, of course, Minnesota has now become a large School of Statistics.

In any case, I had this offer from Stanford. I have to say that at that time, I don't know that I would have accepted anything other than Stanford. I had found Stanford to be really almost an ideal place for someone like myself because I was involved in multivariate, and the offer was joint between Education and Statistics to build up the statistics program in Education. That was really appealing to me. We moved to Stanford in 1961.

The department at Stanford was founded in 1948. Al Bowker was the first chair. There is an interview in *Statistical Science* (Olkin, 1987) that describes the early days and how he became chair. He is a superb manager, an entrepreneur for statistics. Shortly before I came, Herb Solomon became chair, because Al Bowker became Dean of the Graduate School, and Herb was actually the one who hired me. Herb was also a vigorous supporter of statistics. The faculty consisted of Solomon, Herman Chernoff, Charles Stein, Jerry Lieberman, Manny Parzen, Rupert Miller, Lincoln Moses, Vernon Johns and Sam Karlin. I believe I haven't omitted anyone. I was the next there and Kai Lai Chung came the same year I did. Now a critical point in Bowker's thinking was that Statistics, being a small department, would never have too large a faculty unless it had joint appointments, which would mean two people in every billet. But furthermore, he firmly believed that statistics should have all these tentacles and connections, and I think that came out of his being at the statistical research group at Columbia, where once again Hotelling was one of the leading lights. Before too long, we had joint appointments with myself in Education, Ted Anderson in Economics, Tom Cover with Electrical Engineering, Karlin with Mathematics, and Moses with the medical school. The Stanford of the early 60s was not only an exciting place, it was a phenomenally cohesive type of place. At lunch, there was a game of hearts that was very cut-throat and had mathematicians and statisticians. There was a bridge game, there was a go game, and some people went swimming. In 1961, Statistics did not have its own building, nor did Mathematics. Math was scattered over campus, so that some of the mathematicians were in the same building as Statistics. In 1964, the mathematicians got their own building, and Statistics had Sequoia Hall all to its own. But early on, it did not.

**Sampson:** Ingram, it is my impression that Stanford has changed over time in terms of the social relationships in the department?

**Olkin:** Well, I think that's definitely true. You have to remember that in the 60s, most of the faculty lived on campus, and so we were not only close geographically to the Statistics Department, we were also close to one another. The demographics have changed in the Department and the housing has changed. I don't know where everyone lives now, but some faculty might live in Redwood City or San Francisco, and some might live closer to San Jose.

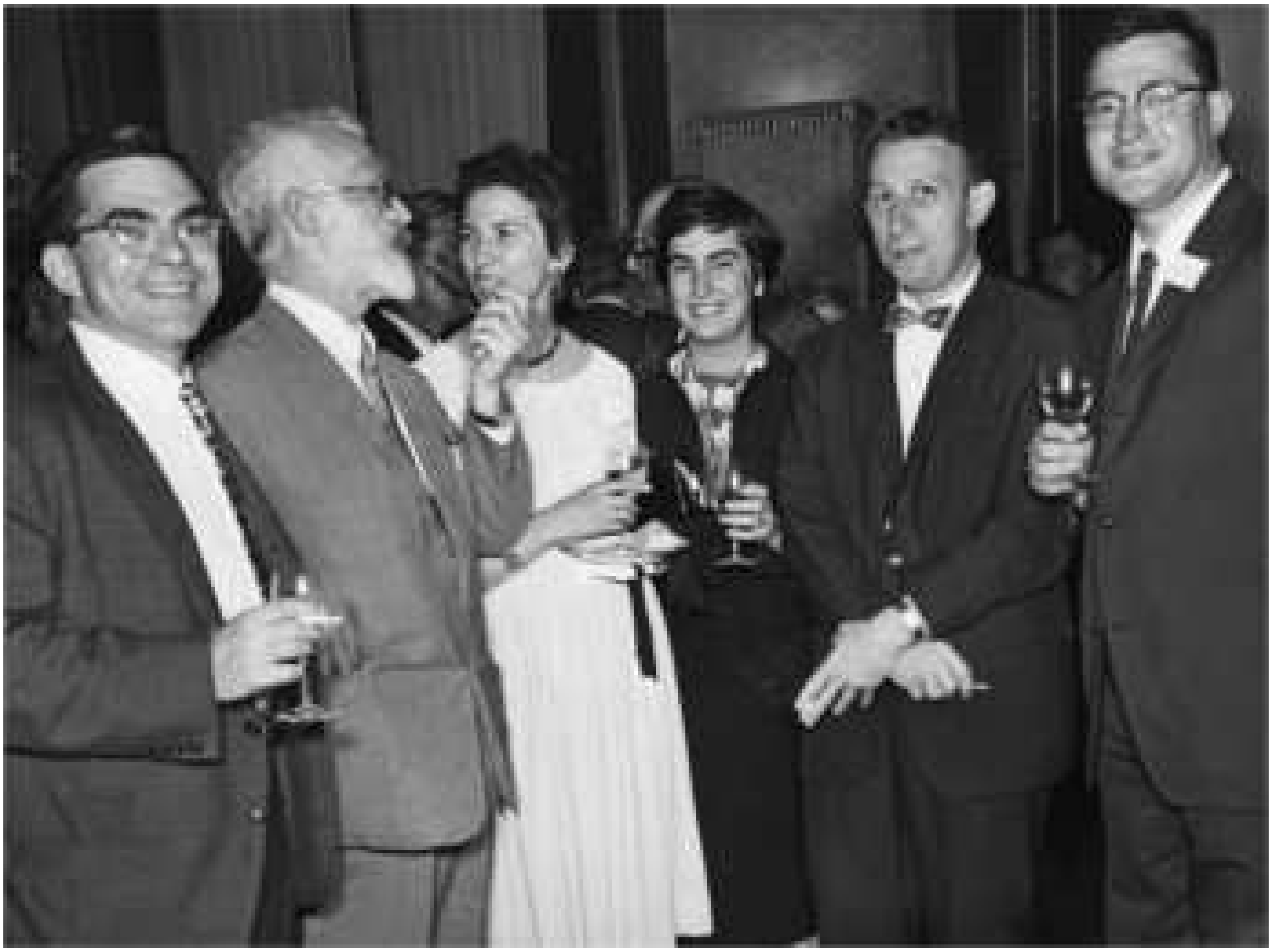

FIG. 10. *Sam Greenhouse, R. A. Fisher, Unknown, Carol Parzen, Ingram Olkin, Manny Parzen (l-r), Paris, 1961. ISI Meeting.*

Also there are now a lot more pressures, a lot more two career families than there were in the 1960s, and as a result of this, people are spending more time away from the office, whereas in the early days the office was sort of "Grand Central Station" in many ways. All this makes for different interactions. People come in and do their work, so to speak, and then maybe go home. The computer has certainly facilitated all of this. There are more closed doors and there's certainly less interaction. My conversation with people in other universities is that the same is true everywhere. People can do an entire job away from the main office. It would be interesting to see a study across universities' departments in which we counted the number of joint publications among faculty who were in the same department, and see how that has changed over time.

**Sampson:** Stanford has always been a wonderful beacon of academic statistics. Do you think it's maintained that role and kept its illustriousness?

**Olkin:** I've thought a lot about how do great departments maintain themselves, and why do some great departments go down and then other departments, who are not that well-known, suddenly become well-known. An important ingredient is the type of young people you bring in. Of course you have to bring in very good people if the department is to maintain its stature. But you also need people who interact with the profession, who are not isolates, because you want the department to in some way become a "domain of attraction." I think that Stanford has been very, very fortunate in being able to attract a number of young people who are clearly not only excellent, but will keep up the stature of the Department. As I look at other places, I think that we have been among the fortunate in the quality of our young people. There are several other departments that I think have also been successful in this. For instance, I think Carnegie–Mellon has done an excellent job in being able to replace itself. Whether a department will be able to maintain its stature is a continuing problem.

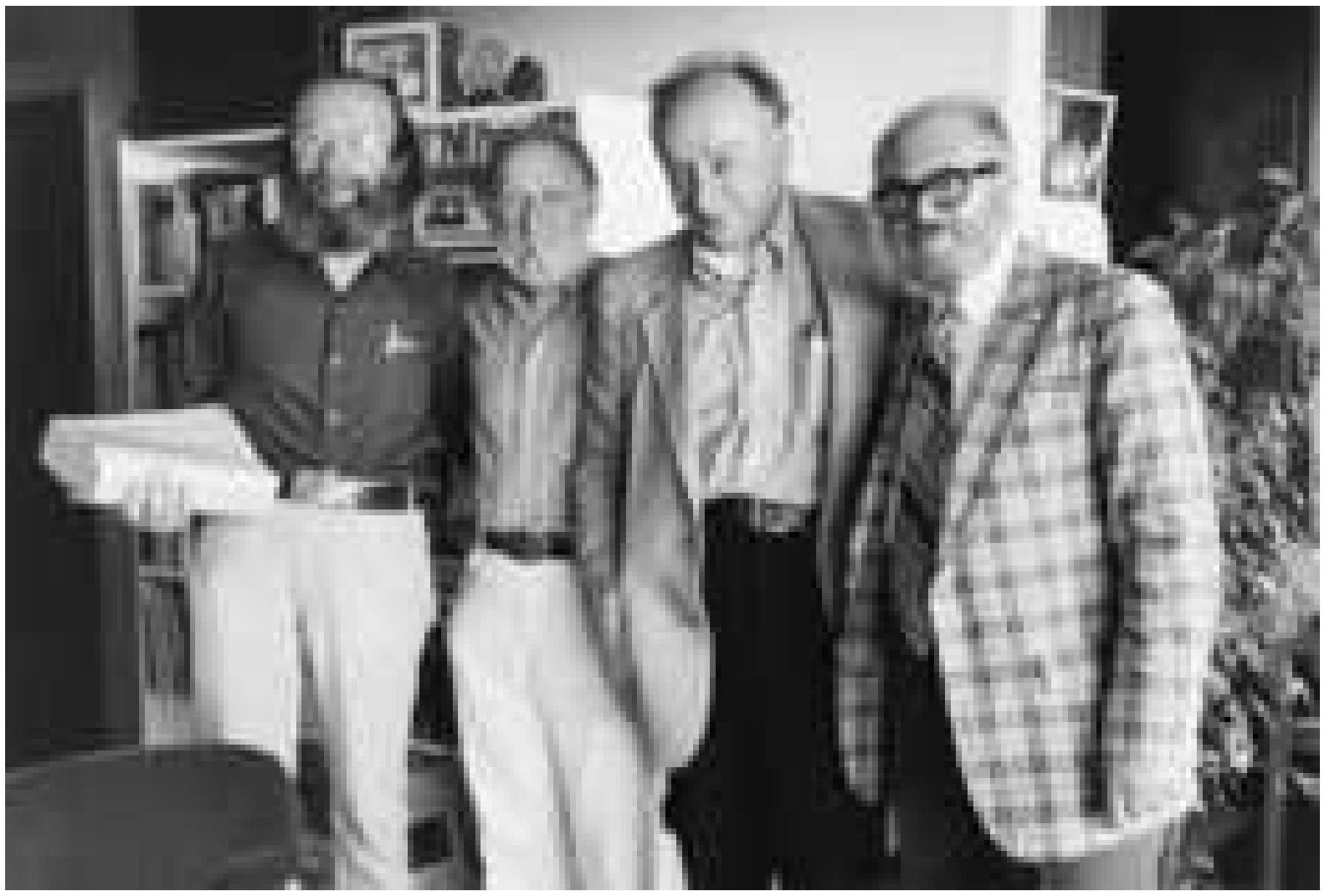

FIG. 11. *Jack Kiefer, Ingram Olkin, Milton Sobel, Unknown, 1981.*

**Sampson:** Ingram, can you talk about what you've seen in terms of the change in the School of Education at Stanford during your time there?

**Olkin:** Now, 1961 was a time when the School of Education was in the throes of trying to become more research oriented, whereas previously they had been oriented towards the practice of education. I was hired at the same time as Dick Atkinson who was half in Psychology and half in Education. Shortly af-ter I came, Janet Elashoff and Lee Cronbach joined the faculty and we formed what you might call an educational statistics group similar to what a biostatistics group would be. We had seminars and we had a Ph.D. program. One of the nicest times in my association with the School of Education was when Rosedith Sitgreaves was a colleague. Sitgreaves, Cronbach and I formed a very nice trio, and quite a number of students graduated in educational statistics. What I want to note is that aside from the people and what they were doing in the program, one of the advantages of being associated with the School of Education is that students had problems that were somewhat different, and it was an applied area that posed many new problems, mostly in multivariate analysis. It was also the genesis of meta-analysis, so I value that association tremendously.

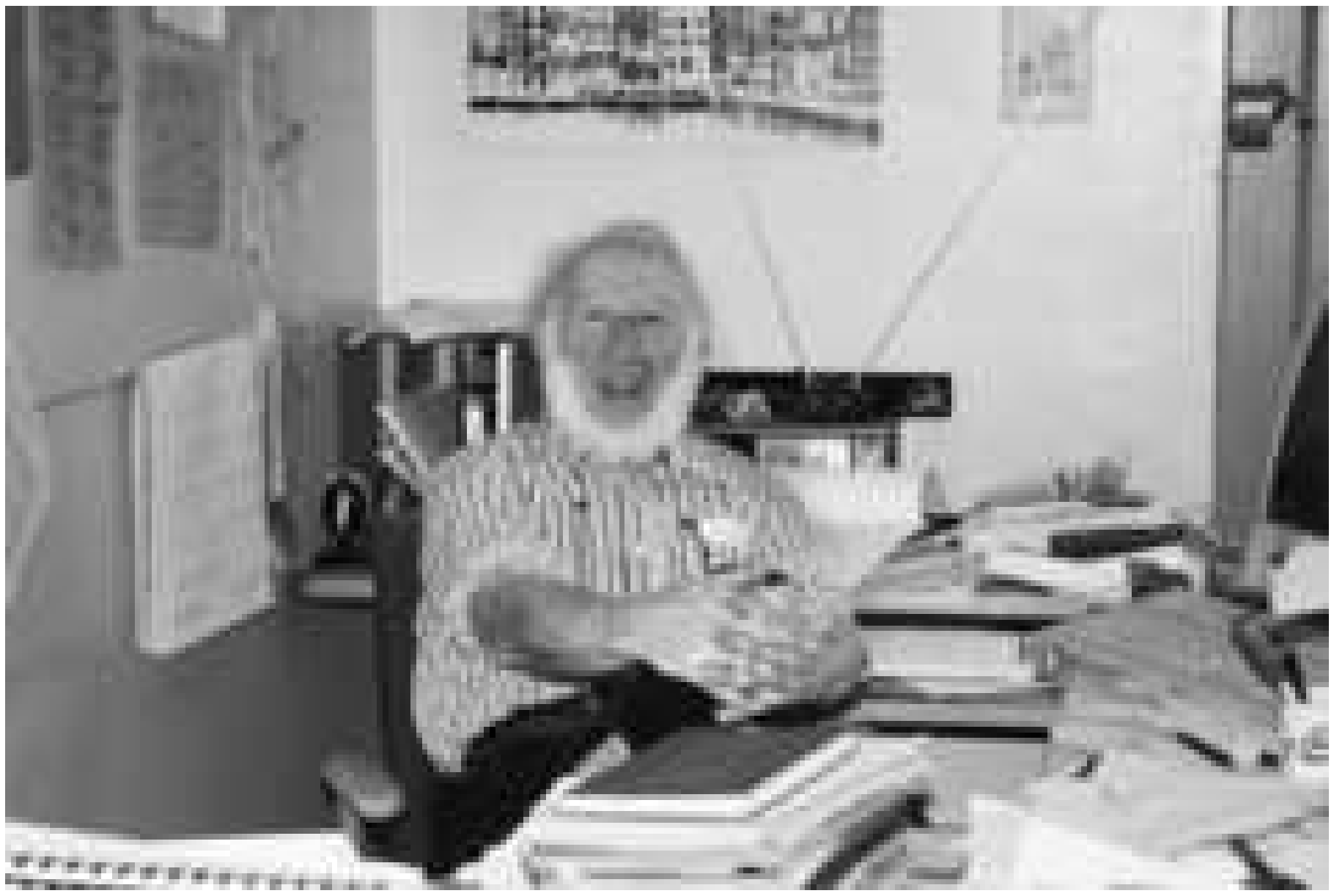

FIG. 12. *Ingram Olkin in his Sequoia Hall office, Stanford University, Spring, 1996.*

**Sampson:** Ingram, do you want to mention your closer friends on the faculty at Stanford, for instance, Jerry Lieberman and Herb Solomon?

**Olkin:** When I first went there it was clear that Jerry and I would become close friends. And we maintained that friendship until his death a number of years ago. I was very close with him. I was also very friendly with Herb Solomon from way back, and he was also a close friend. These were friendships that were maintained over a very long period of time.

**Sampson:** I know that when Jerry was losing his strength and his ability to communicate because of having ALS, you were one of his faithful visitors. You were there several days a week to spend time with him. That must have been both rewarding and difficult.

**Olkin:** Jerry had the ability, even though he could barely speak, to make people feel welcome and want to visit. It was never a depressing time until really close to the end. But it was a hard time because it was not clear how much he understood after a while. Early on, he was quite lucid and was able to communicate. But ALS is such a debilitating disease, so that after a certain point one has to concentrate just on surviving. It's very difficult to think about any other aspects.

**Sampson:** I remember visiting you during that time and going with you to visit Jerry. He had lost totally his ability for speech, but he had a synthetic speech board, and he "talked" with a sense of humor. I remember very vividly he had programmed "take the rest of the day off" and he would hit that as people were leaving.

**Olkin:** Yes, that's absolutely true. He had a number of his favorite comments that were programmed into the synthesizer, and he was always upbeat. He always had a smiling hello, and then we would talk. Basically towards the end, of course, it was a monologue. I would try to bring him up to date on what

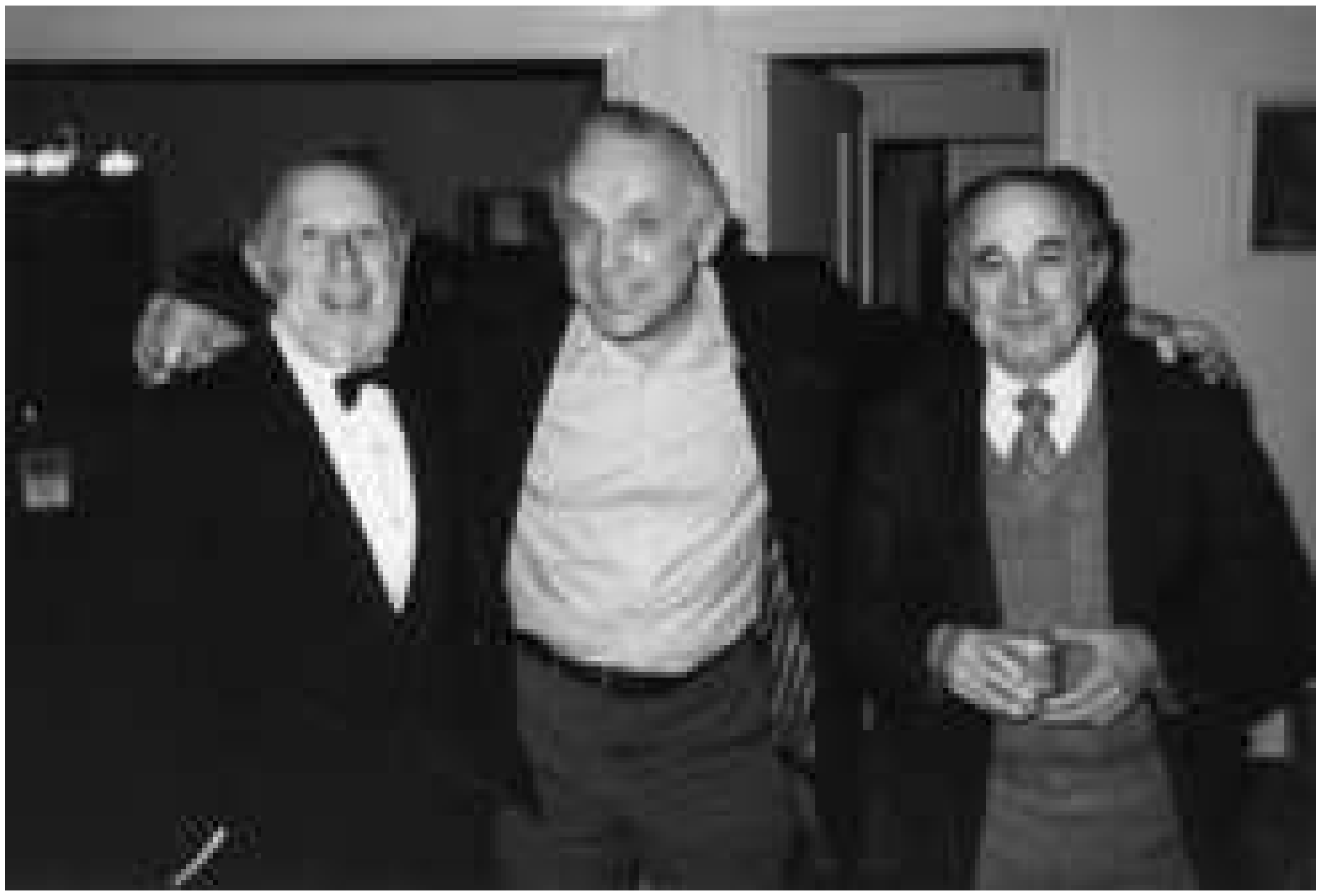

Fig. 13. *Ingram Olkin, Jerry Lieberman and George Resnikoff, 1981.*

was going on, but ALS is a very difficult disease both for the person and for people around.

## STATISTICAL STATESMAN

**Sampson:** The term "statistical statesman" has been applied to you, and I know the people who know you well would agree with this characterization. I'm wondering if you could say what you see your role has been and continues to be as "statistical statesman?"

**Olkin:** There's no question I've been an advocate of statistics. I think at times I've been somewhat forceful and my voice may not have been that welcomed in terms of being an advocate. But I was asked to be on a number of site visits, and that was very good in that it gave me an opportunity to voice to administrations the need for a statistics program or statistics department. I would guess that I was on ten or fifteen such site visits during my lifetime. At a place like the University of Michigan, we were able to make some suggestions which I think, subsequently, led to their making offers and building the department. At Mt. Sinai Hospital, I think it's almost true to say that I helped save one of the statistical programs. This also has occurred abroad and, in fact, I just came back from Croatia where they're talking about starting a doctoral program in statistics. This is within a mathematics department, and they do not have any sense of the role of statistics in applications, and it was gratifying to be able to illuminate them on a lot of these issues. So I think that I've had an effect in that respect.

**Sampson:** There are some materials I've seen from which I have taken this direct quote. Someone described you as a "tireless campaigner for improving and increasing opportunities in statistics for women."

**Olkin:** That's a nice statement, and I really appreciate it. I hope I deserve it. Let me tell you a little bit of how it all started. I think it's apparent to most people that there are very few tenured women in departments of statistics. There are now more tenured women in biostatistics.

But throughout history there were not many tenured women in academia. And what I noticed was that in many departments, especially in small departments, or in departments where a woman would be the lone statistician in the math department, when it came time for tenure, it would be very difficult to make an assessment. The mathematicians would evaluate applied work very differently from theoretical work. It would be very hard for this lone woman to be able to get letters of recommendation. I had the idea that if we could invite some women in their fourth and fifth year towards tenure to Stanford for two summers, that it would give them a chance to broaden their scope, write-up their results and, hopefully, get involved with some of the faculty. The optimal outcome of these visits would be that one of the Stanford faculty would know this person sufficiently well to be able to write a letter of recommendation. And certainly this opened doors for some of these women.

**Sampson:** When did you start this?

**Olkin:** It's now at least 15 years.

**Sampson:** You started it on your own NSF grant?

**Olkin:** The way it started is that I proposed this to the National Science Foundation. They said, "well, we'll try this" and they gave me a supplement to my multivariate grant. I was able then to go to the Associate Provost and make the case that this was a unique program in the United States, that it would help women, and that Stanford could be part of this by waiving overhead which would permit us to invite more women with the same amount of money. Stanford did waive the overhead.

**Sampson:** How many women have come in the years that you've been doing this?

**Olkin:** Well, we used to have somewhere between 1 and 4 women each summer. Over a 15 year period, there were probably 15 to 20 women who came through the program, and I will say most of them did get tenured. I think it just was an inspiration to them. It also looked good on their vita to be able to say that they were invited under an NSF program to be at Stanford.

**Sampson:** Were you involved in other ways with mentoring women?

**Olkin:** I had a number of female Ph.D. students and, in general, I was able to advise them on what would be in store for them after graduation, and to try to help them in deciding on job offers. From many women, not only from Stanford, I got phone calls trying to discuss different offers, what was positive and negative, and I would try to give them some unbiased advice. Not from my point of view, but more from their point of view and what their needs were. It turned out that this is a contagious process. If you help one person, the word gets around, and then you help another one, and another one, and before you know it, you're sort of a central agency for giving advice.

**Sampson:** It is to your tribute that you were the first male to win the Elizabeth Scott Award. That must have been gratifying to you.

**Olkin:** It was really gratifying. Of course, you have to remember I'm the father of three daughters. I became aware of gender bias in science very early on when we were in England. Our oldest daughter was in school, and she wanted to take some science courses and they wanted to counsel her out of it. Well, I found myself being an advocate in that direction and insisting on this not happening. And this happened over and over again in different ways. So I was really aware of the problem. I've also talked to a lot of women who have Ph.D.s and I've discovered that often there was a parent involved who stated that their daughter was not to be discriminated against and fought for them. We now know that many women are counseled out of the sciences, and it does take an advocate.

**Sampson:** I know that you have had a number of other honors. You received an honorary degree from DeMontfort University in England; also a CCNY Distinguished Alumnus Award, an ASA Founders Award, a Wilks Medal, a Lifetime Contribution Award from the American Psychological Association, and a Fisher Lectureship, among others. I'm wondering, are there any favorites among these for you?

**Olkin:** Certainly the Elizabeth Scott Award, but the other one that I really valued a lot was the Wilks Medal. Wilks was one of my heroes, in part, because he was really a statistical statesman. He fought in many ways for the furtherance of statistics, and even though I was not at Princeton, I knew about this and on occasion discussed this with him. Wilks was a theoretician, he had students at Princeton, he was an editor, and he was involved in helping the profession. In many ways I thought of myself trying to follow in the various directions that he had. So that award was really a very pleasing one.

**Sampson:** Would you call him a role model?

**Olkin:** In many ways he was a role model. Hotelling was a different kind of person and fought different kinds of battles, but the battles that Wilks fought were very similar to the battles that I as person could see myself doing. And so, yes, Wilks was definitely a model for me.

## FAMILY LIFE

**Sampson:** We've been talking a lot about your professional life and the fullness of your professional life, but I know you've got a full family life, too. You've been equally involved with Anita and your three daughters. I'm curious how you met Anita. I don't think I ever heard that story.

**Olkin:** What seems to come out in our conversation is that chance played a big role in many aspects of my life, and this was another such occasion. I was in the orchestra at City College. I played trombone. Not very well, I might say, but I played trombone. I played well enough to be in the orchestra, and the orchestra used to play at the basketball games in Madison Square Garden when the City College basketball team was one of the great teams. (They were one of the few teams who won both the NIT and NCAA championships.) My close friend, Andy Gregg, who also played trombone, had a friend by the name of Anita, and Andy introduced me to Anita.

I met her shortly after I joined the service. I was stationed in the east and I was able to meet with her fairly often. Because I was a meteorologist, I was stationed at airports. At that time every plane that took off from an airport had to have the signature of an operations officer and a meteorologist. So I knew about every plane that was taking off. When I had free time and somebody was flying to New York, I would ask the pilot if he could take me, and the answer was often get a parachute and come along. I would all of a sudden come into New York at nine o'clock at night and call Anita and we'd go out. We saw each other for several years this way and also corresponded. Then I was stationed at LaGuardia Airport. I was in New York, actually living at home, and was able to see Anita quite a bit. Next San Francisco contacted LaGuardia and told them that they needed an extra person because they were shorthanded, and would LaGuardia transfer one of its people. But they also had the stipulation that because housing was very difficult in San Francisco, they would prefer if they transferred a single person, and I was chosen to be that person. In any case, I had to go to San Francisco. Well, Anita and I discussed this and we decided to get married. In spite of San Francisco's demand for a single person, we did get married after very short notice. We both took a train to San Francisco, and that was a very happy part of our life.

**Sampson:** You were married right at war's end?

**Olkin:** ` We were married in May of 1945 and the war was coming to an end. We were able to find housing in San Francisco. At the time a lot of people rented rooms. Anita was working and I was working and we would meet afterwards and go to the Officer's Club for dinner, and then just go out. It was a very nice time. I was discharged a year later in 1946, and at that time we returned to New York where I finished City College.

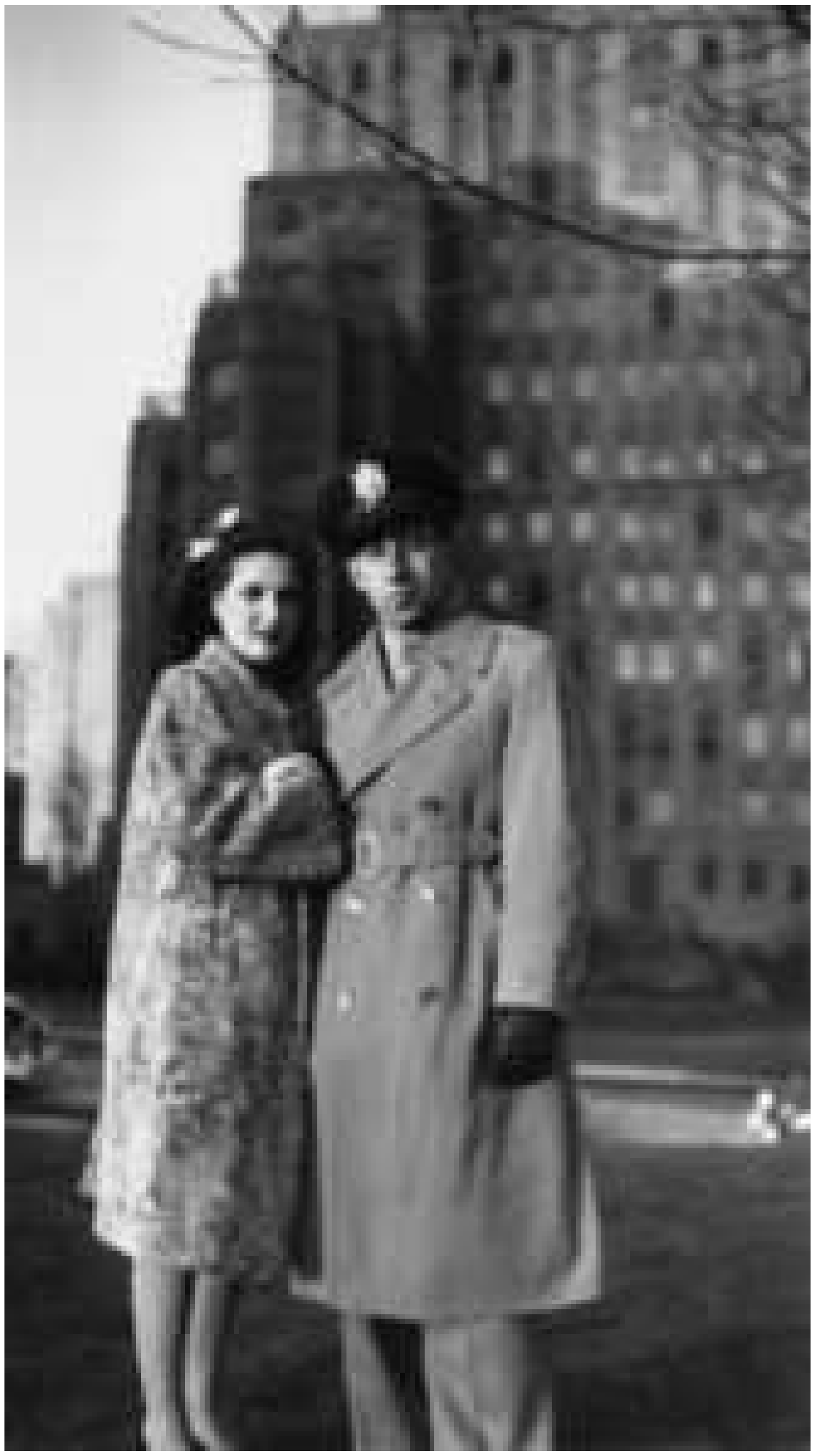

Fig. 14. *Anita and Ingram Olkin, New York, 1945.*

**Sampson:** Vivian, your oldest, was born in North Carolina?

**Olkin:** After Columbia, we went to Chapel Hill, where Vivian was born in 1950. Vivian returned to Chapel Hill about 15 years ago where she now lives with her family.

Then in 1951, we went to Michigan State and Rhoda, my middle daughter was born in 1953 in Lansing. And my youngest Julia is a Stanford child, born in 1959 when I was on sabbatical there.

**Sampson:** Your daughters are very close to you and Anita. I think you spent a lot of time with them when they were growing up.

**Olkin:** Oh yes. I was really very fortunate in having sufficient energy to be able to work during the day and come home in the late afternoons and evenings to spend time with the children. For most young children five o'clock is the "witching" hour and they're tired after the day. At Michigan State, I remember I would come home a bit early and several of us, my colleagues and I, we would take our children and go off to the Cow Barn. Michigan State had an agricultural school, so that it had a lot of areas which were attractive to children. It had a duck pond and cow barns, for example. The children liked the change of scenery. But I had a lot of energy and that's very fortunate when you have children and a work day. After the children went to sleep, I was able to even continue work.

**Sampson:** You use the past-tense "had a lot of energy." I've been at conferences with you recently and at the end of the conference we are all tired. But you are full of energy and tell us "I'm going to have dinner with a grandchild tonight," and you set out for an hour on a bus line somewhere.

**Olkin:** I think those are singular events now-a-days. The ability to work late at night is no longer with me.

**Sampson:** You are a vicious tennis player, and I witnessed that personally. I once heard Harry Joe, who was just half your age when he played you in tennis, say "the only way I know how to beat Ingram"—he felt a bit guilty about this—"was to hit the ball at wide ends of the court and make you just run so hard, that eventually you would wear down." Harry said he could not figure out any other way to beat you.

**Olkin:** Well, I love tennis and I played quite a bit. I was very fortunate always in being able to find a very good doubles partner who could run and help out.

**Sampson:** Harry was describing this in a singles game, Ingram.

**Olkin:** One of my tricks was to wear the wrong socks and what one would call "schlumpy" clothing, and try to psychologically have my opponent not expect very much. That worked up to a certain point, until they discovered what was going on.

[**Laughter**]

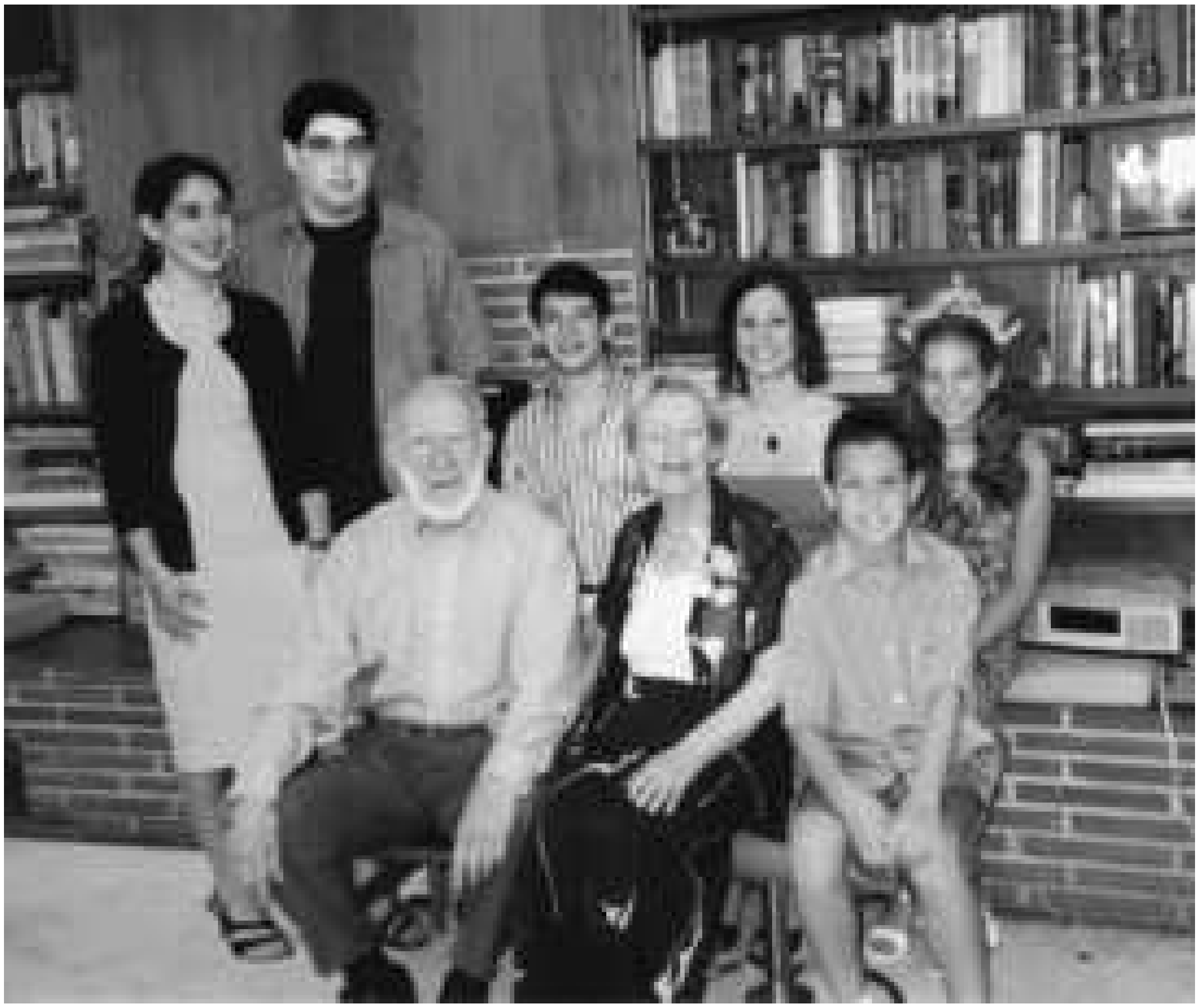

FIG. 15. *Ingram and Anita Olkin and their grandchildren, Leah, Noah, Jered, Sophia, Jeremy and Rachel (l-r), 2005.*

But yes, I did like tennis a lot. I haven't played recently, but I'm going to get back to it.

**Sampson:** You're still swimming.

**Olkin:** Until the recent energy crisis, our swimming pool at the house was heated sufficiently all year round. I really enjoyed getting up in the morning and going swimming for twenty minutes or half an hour everyday. Now I just do it half a year and I have to figure out some way to swim more often.

**Sampson:** Also, my recollection is that you enjoy hiking.

**Olkin:** I have gone on hiking trips—I really love Yosemite. It's been just beautiful up there, and I've gone backpacking, and often with one or two people whom I had known from the hiking groups. I would go, not always with the same groups, and sometimes part of the trip with one, and part with another. There's some beautiful hikes in Yosemite. I would go up to the high country and then hike for about a week, and that was just absolutely magnificent. Your could hike, and then all of a sudden there's an opening and the vista is just breathtaking. I really enjoyed that, and I did that regularly for many years.

## PERSPECTIVES

**Sampson:** We are coming to the end of our conversation. And there a couple of things yet I'd like to you touch upon. The first is to ask if there is any advice you'd like to offer young statisticians who are just starting their careers.

**Olkin:** I think that there are a number of professions where the notion of *pro bono* service is intrinsic in the profession. The medical profession is one, as is the legal profession. And I think the statistical profession should be, in many ways, one such. There are a lot of government agencies and panels that discuss important problems relating to society. Most of these involve the analysis of data, and if the statisticians do not participate in this, I think, it would be a serious mistake. Statisticians bring to the understanding of these problems a different orientation, in my opinion, that is not generally the purview of most of the people who are on these panels.

The term *pro bono* is interesting because it really refers to lack of financial compensation. However, it's not *pro bono* intellectually. That is, there's a very big return by being on a lot of these panels.

**Sampson:** Finally I'd like to ask you if you could, and I realize that this is a hard thing to do, look back at your long and fruitful career in statistics, and from this vantage point say what has given you the most satisfaction?

**Olkin:** There are various things that I've done that, I think, in retrospect have been very fulfilling and satisfying. Of course, all of us have a research career, and that's part of it. The multiple collaborations that I have had with you, with Al Marshall, Leon Gleser, Milton Sobel, Larry Hedges, Michael Perlman, Jim Press, and others have been both enjoyable and satisfying. But I think the friends and students whom I have influenced in a positive way have been very satisfying. I've really enjoyed the process of helping a student from beginning to fruition. It is very, very satisfying when students complete a dissertation and go on to live a fruitful life. If they become totally and independently productive, I find that the personal satisfaction is immeasurable.

Also I think my accomplishments in terms of the profession have also been very gratifying and, of course, my role in helping build Stanford into a great statistics department complements all of these. I think this composite was very fulfilling and really makes me feel satisfied when I look back at my career.

**Sampson:** Thank you Ingram. That's an inspiration for all of us, and thank you for this interview.

## ACKNOWLEDGMENTS

We wish to thank Diane Hall who did a superb job, as always, of transcribing the original tapes. Also we want to thank Leon Gleser for his advice prior to the conversation and his assistance in editing the print version of this conversation.

## REFERENCES

ABRAMOWITZ, M. and STEGUN, I., eds. (1972). *Handbook of Mathematical Functions with Formulas, Graphs, and Mathematical Tables.* National Bureau of Standards Applied Mathematics Series **55**. US Government Printing Office, Washington, DC. MR0167642

ACZÉL, J. (1966). *Lectures on Functional Equations and Their Applications.* Academic Press, New York. MR0208210

CRAMÉR, H. (1946). *Mathematical Methods of Statistics.* Princeton Univ. Press, Princeton. MR0016588

DAS GUPTA, S., EATON, M. L., OLKIN, I., PERLMAN, M., SAVAGE, L. J. and SOBEL, M. (1972). Inequalities on the probability content of convex regions for elliptically contoured distribution. *Proc. Sixth Berkeley Symp. Math. Statist. Probab.* **2** 241–265. Univ. California Press, Berkeley. MR0413364

DEEMER, W. and OLKIN, I. (1951). The Jacobians of certain matrix transformations useful in multivariate analysis. *Biometrika* **38** 345–367. MR0047300

FELLER, W. (1950). *Introduction to Probability Theory and Its Applications* **1**. Wiley, New York. MR0038583

GLASS, G. V., MCGRAW, B. and SMITH, M. L. (1981). *Meta-analysis in Social Research.* Sage, Publications, Beverly Hills, CA.

GLESER, L. J., PERLMAN, M. D., PRESS, S. J. and SAMPSON, A. R. (1989). *Contributions to Probability and Statistics*: *Essays in Honor of Ingram Olkin.* Springer, New York. [Biography reprinted in *Linear Algebra and Its Applications* 1–15 (1994). MR1274406]. MR1024318

HEDGES, L. V. and OLKIN, I. (1985). *Statistical Methods for Meta-Analysis.* Academic Press, New York. MR0798597

HOEFFDING, W. (1940). Masstabinvariante korrelationstheorie. *Schriften des Mathematischen Instituts und des Instituts fur Angewandte Mathematik der Universitat Berlin* **5** 179–233.

JONES, L. V. and OLKIN, I. (2004). *The Nations Report Card*: *Evolution and Perspectives.* Phi Delta Kappa, Bloomington, IN.

KENDALL, M. G. (1944). *The Advanced Theory of Statistics.* **1**. J. B. Lippincott Co., Philadelphia. MR0010934

KENDALL, M. G. (1946). *The Advanced Theory of Statistics.* **2**. Charles Griffin and Company, Ltd., London. MR0019869

MARSHALL, A. and OLKIN, I. (1967). A multivariate exponential distribution. *J. Amer. Statist. Assoc.* **62** 30–44. MR0215400

MARSHALL, A. and OLKIN, I. (1979). *Inequalities: Theory of Majorization and Its Applications.* Academic Press, New York. MR0552278

OLKIN, I. (1987). A conversation with Albert H. Bowker. *Statist. Sci.* **2** 472–483. MR0933739

OLKIN, I. (1997). *The History and Development of Meta-Analysis (DS041) and A Conversation with Ingram Olkin (DS042). Distinguished Lecture Videotapes.* American Statistical Association, Washington, DC.

OLKIN, I., GHURYE, S. G., HOEFFDING, W., MADOW, W. G. and MANN, H. B., eds. (1960). *Contributions to Probability and Statistics*: *Essays in Honor of Harold Hotelling.* Stanford Univ. Press, Stanford, CA. MR0120692

OLKIN, I. and SACKS, J. (1988). *Cross-Disciplinary Research in the Statistical Sciences.* (Report of a Panel of the Institute of Mathematical Statistics.) IMS, Hayward, CA.

PRATT, J. W. and OLKIN, I. (1958). On a multivariate Tchebycheff inequality. *Ann. Math. Statist.* **29** 226–234. MR0093865